\documentclass{article}
\usepackage{arxiv}
\usepackage{graphicx,amsmath,amsfonts,amscd,amsthm,amssymb,pstricks}
\usepackage[UKenglish]{babel}
\usepackage{color,natbib}
\usepackage{hyperref}
\usepackage{tikz}
\usepackage{subfigure}
\usepackage{authblk}

\title{Machine learning for beam dynamics studies at the CERN Large Hadron Collider}

\author[1]{P.~Arpaia}
\author[2]{G.~Azzopardi}
\author[3]{F.~Blanc}
\author[4]{G.~Bregliozzi}
\author[2]{X.~Buffat}
\author[3,2]{L.~Coyle}
\author[5,2]{E.~Fol}
\author[1,2]{F.~Giordano}
\author[2]{M.~Giovannozzi\thanks{Corresponding author: massimo.giovannozzi@cern.ch}}
\author[3,2]{T.~Pieloni}
\author[1]{R.~Prevete}
\author[2]{S.~Redaelli}
\author[2]{B.~Salvachua}
\author[2]{B.~Salvant}
\author[3]{M.~Schenk}
\author[2]{M.~Solfaroli~Camillocci}
\author[2]{R.~Tom\'as}
\author[6]{G.~Valentino}
\author[6,2]{F.F.~Van~der~Veken}
\author[2]{J.~Wenninger}
\affil[1]{Dipartimento di Ingegneria Elettrica e Tecnologie dell'Informazione (DIETI), Universit\`a degli studi di Napoli Federico II, 80125 Napoli, Italy}
\affil[2]{Beams Department, CERN, Esplanade des Particules 1, 1211 Geneva 23, Switzerland}
\affil[3]{Ecole Polytechnique Federale Lausanne, 1015, Lausanne, Switzerland}
\affil[4]{Technology Department, CERN, Esplanade des Particules 1, 1211 Geneva 23, Switzerland}
\affil[5]{Johann Wolfgang Goethe Universit\"at, Max-von-Laue-Str. 9, 60438 Frankfurt, Germany}
\affil[6]{University of Malta MSD2080, Msida, Malta}

\begin{document}
\maketitle
\begin{abstract}
Machine learning entails a broad range of techniques that have been widely used in Science and Engineering since decades. High-energy physics has also profited from the power of these tools for advanced analysis of colliders data. It is only up until recently that Machine Learning has started to be applied successfully in the domain of Accelerator Physics, which is testified by intense efforts deployed in this domain by several laboratories worldwide. This is also the case of CERN, where recently focused efforts have been devoted to the application of Machine Learning techniques to beam dynamics studies at the Large Hadron Collider (LHC). This implies a wide spectrum of applications from beam measurements and machine performance optimisation to analysis of numerical data from tracking simulations of non-linear beam dynamics. In this paper, the LHC-related applications that are currently pursued are presented and discussed in detail, paying also attention to future developments.
\end{abstract}
\keywords{Machine Learning, Beam dynamics, LHC}
%
%

\section{Introduction}

Machine Learning (ML) is the process of building a mathematical model based on sample data, known as ``training data'', in order to make predictions or decisions without being explicitly programmed~\cite{Mitchell97}. ML is a subset of Artificial Intelligence (AI), and encompasses a number of learning paradigms, including Supervised Learning (SL), Unsupervised Learning (UL), and Reinforcement Learning. Typical ML tasks include classification, regression, clustering, anomaly detection, dimensionality reduction, and reward maximisation~\cite{bishopml}. 

The process involved in using ML to train a mathematical model to achieve successfully a particular task involves a number of steps. These include data collection and curation, feature (input) engineering, feature selection and dimensionality reduction, model hyper-parameter optimisation, model training, performance evaluation and finally deployment in operation.

In the SL paradigm, ML algorithms are trained on labelled data sets, meaning that there exists a ground-truth output (continuous or discrete) for each input. On the other hand, in UL~\cite{Bonaccorso:2671442} no ground-truth output is available, and the algorithms typically attempt to discover patterns in the data. Reinforcement Learning~\cite{Sutton1998} is more suitable for sequential decision-making tasks, where the goal is to maximise the cumulative reward.

Machine Learning has recently enjoyed success in a number of applications, from computer vision~\cite{8624570} to bioinformatics~\cite{doi:10.1002/prot.25832}, stock market prediction~\cite{8456512}, fraud detection~\cite{8038008}, robotics~\cite{4154938}, and transportation~\cite{8325485}. This has come about due to a number of factors, such as an explosion in Big Data, advances in computational power (in particular the use of Graphics Processing Units), and also the development of more sophisticated ML techniques such as deep learning~\cite{deeplearning}, an end-to-end learning process that requires large amounts of data and automates certain manual parts of the classical ML process, such as feature engineering and extraction.

In recent years, ML techniques have found their way into the field of accelerator physics. The first attempts to use these techniques in beam diagnostics and beam control systems already date from a few decades ago~\cite{Bozoki:1994xm,LeBlanc:2012zz}, but only recently some sizeable progress has been made (see, e.g.~\cite{ML1,ML2,ML3,ML4,ML5,ML6} and references therein, for a sample of recent applications of ML to Accelerator Physics topics). The Accelerator Physics community is globally recognising the need and usefulness of ML techniques, which has resulted in a growing number of conferences and workshops and the publication of a white paper with a detailed review of the state-of-the-art and several recommendations to encourage uptake of such techniques in Accelerator Physics laboratories~\cite{Edelen:2018jid}. 

At the CERN Large Hadron Collider (LHC), several ML applications are under study and are actively pursued to assess the potential benefits. Generally speaking, given the intrinsic complexity of the LHC in terms of number of operational systems, amount of data collected and made available for on-line or off-line analyses, variety of beam dynamics configurations, such as optical configurations, and beam dynamics phenomena,  single-particle, collective effects (coherent and incoherent), it is clear that all this is an ideal playground where case studies for ML applications can be found. Note that the results of numerical simulations can also provide an interesting ground for applications of ML to the data analyses. In this paper, the most interesting applications devised so far will be presented and discussed in detail. This is also in view of applications for potential future machines, such as the Future Circular Collider (FCC)~\cite{Benedikt:2651300}.

The plan of the paper is the following: in Section~\ref{sec: LHC_and_OMC} a brief description of the LHC ring is presented, including also some highlights of the ML activities linked with the linear optics measurements and correction, which are among the first LHC activities that explored ML applications. In Section~\ref{sec: coll} the application of ML to the beam commissioning of the LHC collimation system is presented. In Section~\ref{sec: lifetime}, ML is applied to the topic of improving machine performance, in terms of particle losses and beam lifetimes. In Section~\ref{sec: instab} the detection of coherent beam instabilities and their automatic selection by means of ML is presented. In Section~\ref{sec: heating}, ML is used to classify pressure readings and transform them into estimates of potential heating effects. Finally, in Section~\ref{sec: DA} the application of ML to the analyses of tracking data from beam dynamics simulations is discussed. Finally, conclusions and outlooks to the future are presented in Section~\ref{sec: conc}. 
\section{LHC ring and optics measurements and correction}\label{sec: LHC_and_OMC}
\begin{figure}[ht]
  \centering
  \includegraphics[width=0.9\linewidth,clip=]{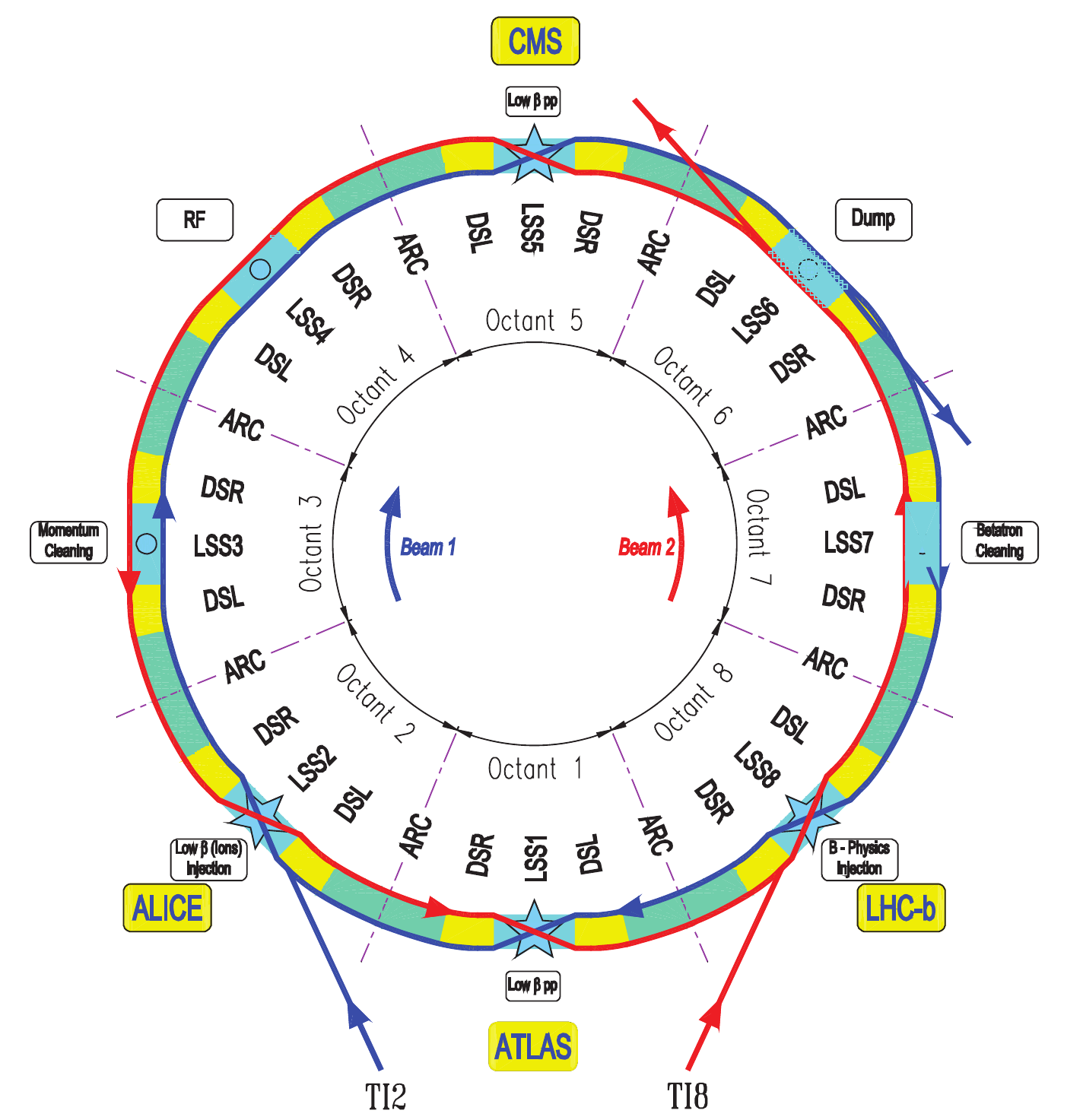}
  \caption{Layout of the LHC ring}
  \label{fig:LHClayout}
\end{figure}

The layout of the LHC ring is shown schematically in Fig.~\ref{fig:LHClayout} (see~\cite{Bruning:782076} for more detail).
In the Run~2 (2015-2018), the LHC accelerated proton and ion beams from an energy of $450$~Z~GeV to a maximum flat-top energy of $6.5$~Z~TeV where most of the physics programme was performed (see, e.g.~\cite{Run2}).

The eight-fold symmetry is clearly visible as well as the main function of each long straight section. Note that the RF system is located in the same straight section as the non-distributed beam instrumentation devices, such as transverse and longitudinal profile monitors and beam current monitors. It is also worth mentioning that LHC Sectors are defined as the ring parts in between the mid-points of consecutive Octants, e.g. Sector~4-5 is defined as the fraction of the ring circumference between the mid-point of Octant~4 and the mid-point of Octant~5. 

Great emphasis has been put on the design and development~\cite{ATS} of the linear optics and on its measurement and correction and these efforts have been rewarded by outstanding results~\cite{OMC1,OMC2,OMC3,OMC4}, which are among the key items that brought to the excellent performance of the LHC. To improve these results even further one should tackle two key issues, namely devise techniques to recognise efficiently faulty Beam Position Monitor (BPM) readings and build effective models to reproduce the impact of field errors distributed in the ring. Both aspects are particularly suited to the use of ML techniques, which have been actively pursued in recent years. 

Optics measurements and corrections at the LHC are incorporating ML techniques in two different forms, namely SL and UL. Supervised methods are used to explore the opportunities to build regression models that aim to reconstruct individual magnet errors from optics perturbations caused by these errors, while currently available correction techniques compute circuits strength settings to compensate the measured optics deviations from design. The preliminary results presented in~\cite{Fol:masterthesis, Fol:WEPAF062, Fol:THPRB077, Fol:FEL19-WEB03} clearly demonstrate the ability of ML-based regression models~\cite{randomforest, lsregression, Ridge, OMP, CNN} to predict the individual quadrupole errors. Furthermore, they potentially improve the quality of optics corrections by providing additional knowledge about the local-error sources in the ring. Recent results obtained from simulations and LHC measurements data are presented in~\cite{regressionmodels}. 

Unsupervised learning, on the other hand, is applied to optics measurements in order to detect faulty BPM signals, which produce nonphysical outliers in the optical functions computed from BPM data. Application of a decision-tree-based Isolation Forest algorithm (IF)~\cite{Liu:2008:IF} significantly improved the quality of the measurements data used to compute the optical functions, avoiding the appearance of outliers and reducing the human effort in data cleaning. 

The general concept of applying anomaly detection techniques to measurements cleaning at the LHC, operational results, and comparison to clustering techniques are presented in~\cite{Fol:WEPGW081}. However, extensive study on cleaning techniques for the LHC optics measurements, recent advances, and future plans are demonstrated in~\cite{faultyBPMs}.
\section{Beam commissioning of the collimation system}\label{sec: coll}
The LHC is susceptible to beam losses from normal and abnormal conditions, which can damage the state of superconductivity in its magnets and eventually lead to a quench~\cite{Bruning:782076}. As a result, the equipment must be protected from any damage or down-time that may be caused due to beam losses.
The LHC relies on a robust collimation system to dispose safely of such unavoidable beam losses. The LHC collimation system consists of around 100 collimators distributed along the 27~km ring, whereby each collimator is made of two parallel absorbing blocks. Each of the four jaw corners can be moved individually using dedicated stepper motors, for a total of about 400 degrees of freedom for the whole system. 
The collimator jaws are positioned with an accuracy of 5~$\mu$m around the circulating beam, with the tightest operational gap being around $1.1$~mm at top energy. 

The halo cleaning performance provided by the LHC collimation system relies on a precise multi-stage transverse setting hierarchy of different collimator ``families'' (primary, secondary, and tertiary collimators; shower absorbers; protection devices)~\cite{bruce15_PRSTAB_betaStar, bruce17_NIM_beta40cm}. The collimator settings are determined following a beam-based alignment (BBA) procedure established in~\cite{assmann:BBA}, to determine the beam centre and beam size at their locations. This procedure moves collimator jaws separately towards the beam halo, whilst monitoring the measured beam loss signal. Each collimator has a dedicated Beam Loss Monitoring (BLM) device positioned immediately downstream, to detect beam losses generated when halo particles impact the collimator jaws. A collimator is said to be aligned when both jaws are centred around the beam after touching the beam halo, which is indicated by a signature spike pattern in the recorded beam loss signal. The local beam size is determined by comparing the aligned jaw positions against a reference beam halo established with the primary collimators~\cite{valentino2012semiautomatic}. At present, the BBA is semi-automated~\cite{valentino2012semiautomatic} and collimation experts are required to manually detect and classify such spikes following training and experience. Given the complexity of the LHC collimation system and of the LHC operational cycle, the procedure to establish settings from injection to collision is tedious and time consuming.

Each year during beam commissioning, the collimators must be aligned to ensure the correct setup for the specific LHC run configuration, prior to achieving nominal operation. They are aligned at different machine states; at injection ($450$~GeV) 79 collimators are aligned, and at flat top ($6.5$~TeV) 75 collimators are aligned. Their settings are monitored along the year, and different collimator setups are required when machine parameters are changed. This alignment procedure is crucial as it is a prerequisite for every machine configuration to set up the system for high-intensity beams. This motivated the development of an automatic method, to allow for collimator alignments to be performed more efficiently and upon request, at regular intervals.

In addition to this motivation, it is noted that collimators are for the moment aligned assuming no tilt between the collimator and the beam, therefore any tank misalignments or beam envelope angles at large-divergence locations could introduce a tilt with respect to the optimum orientation, which might limit the collimation performance. This is a concern in particular if the collimation hierarchy is pushed to tighter retraction between families to optimise the performance~\cite{bruce15_PRSTAB_betaStar}. It is planned to improve the angular accuracy in the future, to optimise further the collimation system performance. A recent study~\cite{Azzopardi:angular} introduced three novel angular alignment methods to determine a collimator’s most optimal angle, however these methods also make use of the semi-automated software and require much longer setup times. 

\begin{figure}[htp]
\centering
\subfigure[\label{fig:yes_spike} Alignment spike - the corresponding collimator is aligned with the beam.]{
\includegraphics[width=0.75\textwidth,clip=]{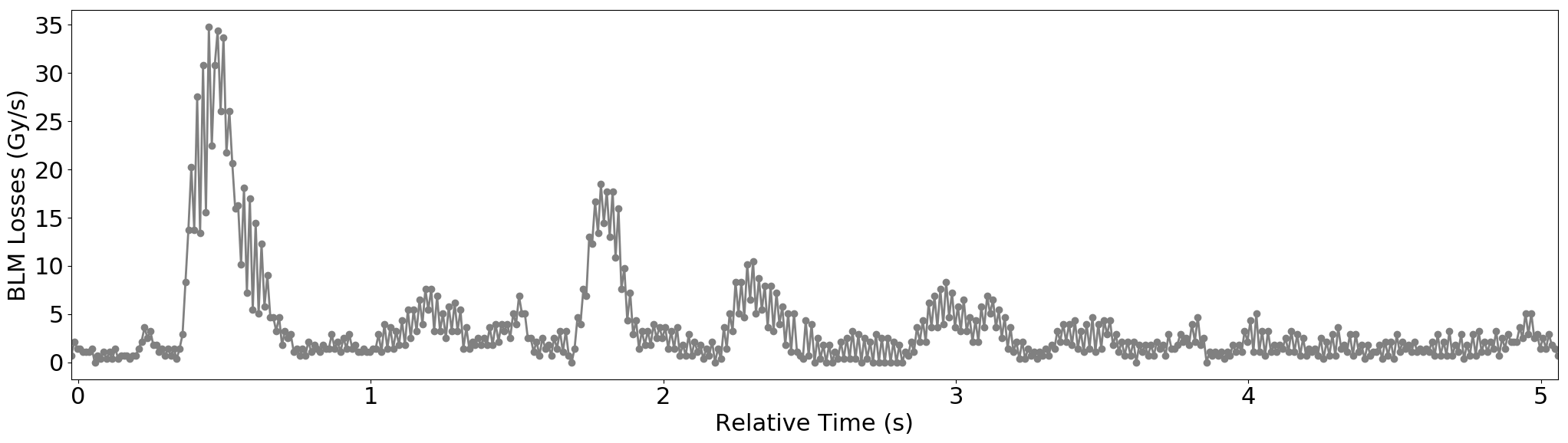}}
\subfigure[\label{fig:no_spike} Non-alignment spikes - the corresponding collimator is far from the beam.]{
\includegraphics[width=0.75\textwidth,clip=]{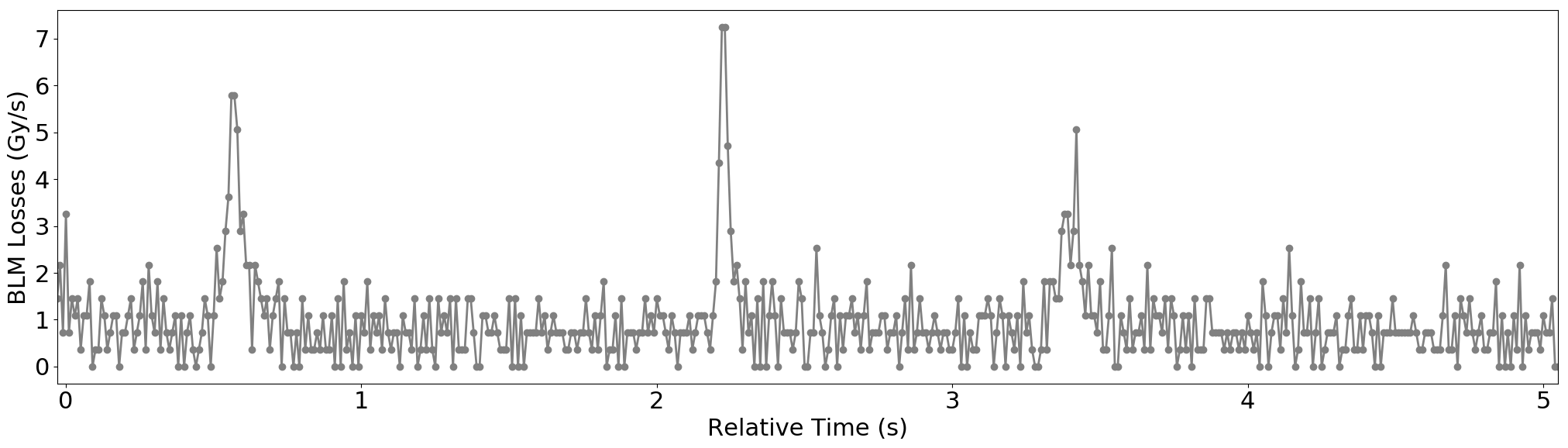}}
\caption{\label{fig:yes_no_spikes} Typical BLM signals at 100 Hz showing (a) a clear alignment spike and (b) non-alignment spikes , after inward collimator movements at approximately t = 0.5 s.}
\end{figure}

Collimator alignment campaigns involve continuously moving the jaws towards the beam, whilst ignoring any non-alignment spikes, until a clear alignment spike is observed. An alignment spike, as shown in Figure~\ref{fig:yes_spike}, indicates that the moving jaw touched the beam halo and is hence in contact with the primary beam. It consists of a steady-state signal before the spike (corresponding to movements of the jaws before the beam is reached), the loss spike itself, the temporal decay of losses, and a steady-state signal after the spike. This second steady-state, with larger losses than the first one, is a result of the continuous scraping of halo particles when the jaw positions are fixed. The further a jaw cuts into the beam halo the more the steady-state signal increases, as the density of the particles near the jaw increases. Any other spikes which do not follow this pattern are classified as non-alignment spikes as shown in Figure~\ref{fig:no_spike}. They do not have a fixed structure and can contain spurious high spikes. Such non-alignment spikes arise due to other factors, i.e.\ beam instabilities or mechanical vibrations of the opposite jaw, thus indicating that the jaw has not yet touched the beam and must resume its alignment. In order to achieve a reliable alignment, one has to be able to correctly identify such alignment spikes. Note that in a single alignment campaign, hundreds of such spikes need to be analysed.

Recent work~\cite{azzopardi:ML}, sought to fully-automate the BBA by automating the process of spike recognition, by casting it as a classification problem, such that ML models were trained to distinguish between the two spike patterns in the BLM losses. Data was gathered from 11 semi-automatic collimator alignment campaigns performed in 2016 and 2018, both at injection and at flat top. A total of 6446 samples were extracted, 4379 positive (alignment spikes) and 2067 negative (non-alignment spikes). The data logged during alignment campaigns consists of the 100 Hz BLM signals and the collimator jaw positions logged at a frequency of 1~Hz. The data extracted for the data set consists of the moments when each collimator jaw(s) stopped moving, when the losses exceeded the threshold defined for the semi-automatic alignment.

Fourteen manually-engineered features were extracted from this data set and were analysed. In order to select the most relevant features, the strength of association between each pair of variables was first analysed using the Spearman correlation. A feature selection analysis was then performed using five different ML models to see how they order the importance of each of the features. The models were individually trained using all features and outputted the features ranked in ascending order, according to their importance. Finally, sequential forward selection algorithm (SFS)~\cite{ref:sfs} was performed, to select the best features with the best hyper-parameters. The SFS algorithm tries all feature combinations by introducing one feature at a time and keeping the best feature for future combinations. 

The resulting five most important features were:
\begin{itemize}
    \item Height (1 feature) - This is calculated by subtracting the average steady state losses before the spike from the maximum value. The average steady state is calculated from the BLM signal after the decay of the previous alignment, until the current collimator was stopped.
    \item Spike decay (3 features) - Exponential fit to the decay in the BLM signal using $ae^{-bx}+c$.
    \item Position in sigma (1 feature) - A beam size invariant way of expressing the fraction of the normally distributed beam interrupted by the jaw, as the beam size in mm varies across locations in the accelerator.
\end{itemize}
These features were used to train and compare six ML models for binary classification, namely; Logistic Regression, Neural Network, Support Vector Machine, Decision Tree, Random Forest, Gradient Boost.

\begin{figure}[ht]
  \centering
  \includegraphics[width=3.5in, trim={2cm 0 2cm 3.04cm},clip=]{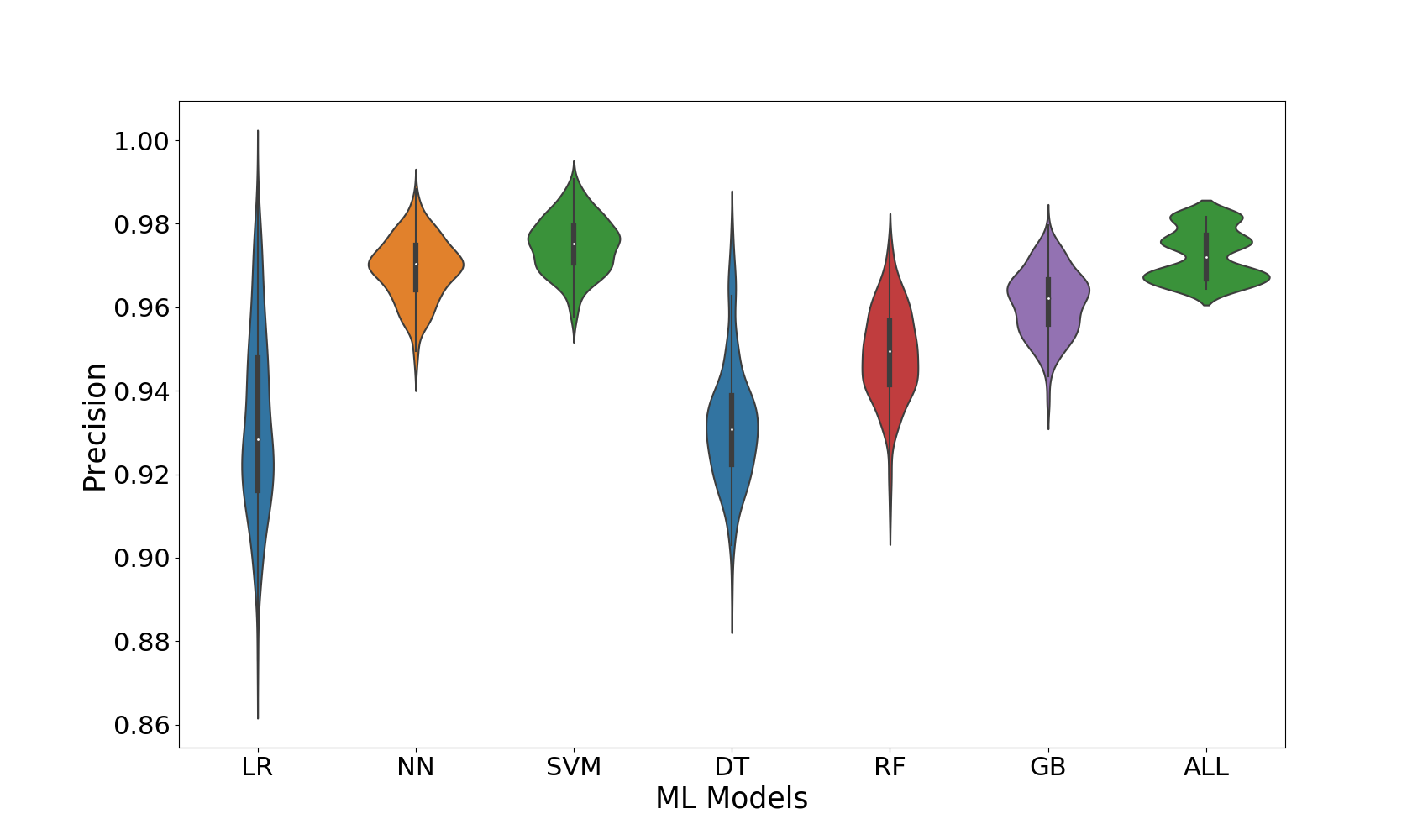}
  \caption{The precision distribution obtained by each model and their Ensemble (LR: Logistic Regression, NN: Neural Network, SVM: Support Vector Machine, DT: Decision Tree, RF: Random Forest, GB: Gradient Boost, ALL: Ensemble).}
  \label{fig:train16+18-test16+18}
\end{figure}

When aligning any collimator it is vital that after the spike detection predicts a collimator to be aligned, then the collimator should have actually touched the beam and is aligned. Otherwise, if the spike detection predicts the collimator to be aligned when the collimator is still far from the beam, this would result in a misalignment and the collimator must be realigned from the beginning. As a result, false detection of an alignment spike is more grievous than not detecting an alignment spike, therefore precision was used as the main performance metric. Each model was analysed in-depth, optimised using hyper-parameters, and thoroughly tested on unseen data. The results were collected by applying a cross-validation on the training set, which was performed by dividing the original training set into ten randomly-defined subsets: nine used for training and the last one for validating the results. This procedure was then repeated 30 times, each time with a different random partition of the training set in order to handle lucky splits. Figure~\ref{fig:train16+18-test16+18} plots the precision distribution obtained by each of the models and their Ensemble. In addition, Tukey’s HSD test~\cite{ref:tukey} was used to determine whether the means of the results obtained by the models were significantly different. These results are included in Figure~\ref{fig:train16+18-test16+18}, such that a different colour implies a significantly different mean.

These results indicate that the Support Vector Machine obtained the best precision, with a mean similar to the Ensemble, which has a more stable range of results. The Ensemble model was incorporated into the BBA software~\cite{Azzopardi:architecture}, together with the necessary threshold-selection algorithm~\cite{Azzopardi:threshold} and cross-talk analysis~\cite{azzopardi:crosstalk}, to transform the semi-automatic alignment into a fully-automatic one. 

\begin{figure}[ht]
\centering
	\includegraphics[width=3.0in,clip=]{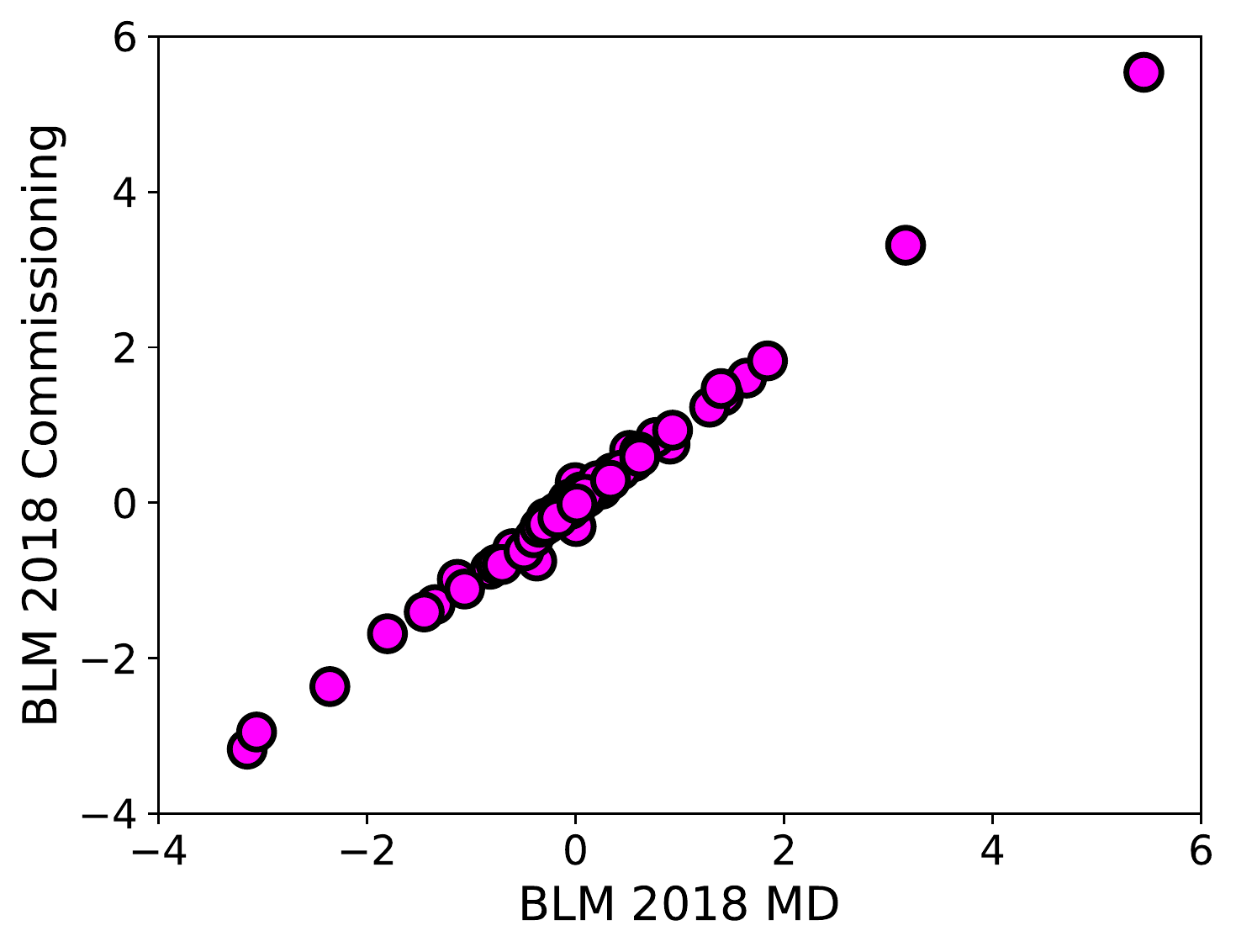}
	\caption{\label{fig:results_parallel} Parallel alignment MD results at injection: the beam centres obtained in the 2018 commissioning using BLMs as a function of the centre values obtained in MD. The results show a good correlation.}
\end{figure}

This new fully-automatic alignment software was successfully used throughout 2018 LHC operation. The first version was used during commissioning, such that the collimators in the two beams were automatically aligned sequentially, at injection and flat top. A machine development (MD) study was then scheduled to test the alignment of the collimators of the two beams in parallel. Finally, another MD was scheduled to test the parallel fully-automatic software with angular alignments.

The collimator centres measured at injection with BLM detectors during injection commissioning in 2018 are similar to the centres obtained during the parallel alignment MD. This is depicted in Figure~\ref{fig:results_parallel}, evidently showing the reproducibility of the LHC and the quality of orbit and optics correction that enable such an excellent stability of the collimator alignment.

\begin{figure}[!ht]
\centering
	\includegraphics[width=3.5in,clip=]{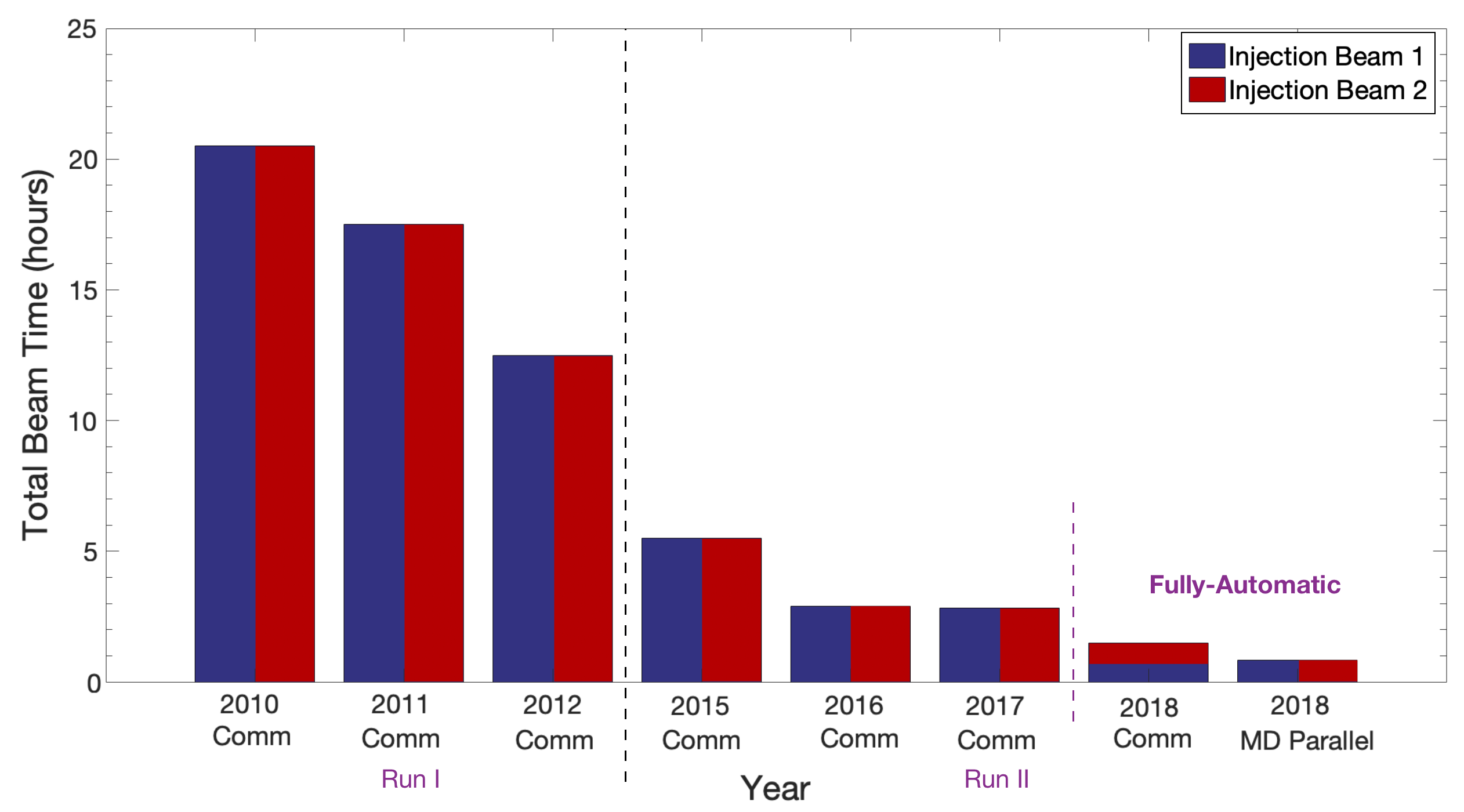}
	\caption{\label{fig:align_time_par} The time to align the collimators at injection commissioning, compared to the 2018 parallel alignment MD, from~\cite{azzopardi:results}.}
\end{figure}

The time to align the collimators at injection was decreased by 71.4\%, compared to the semi-automatic alignment in 2017, namely from 2.8 hours to 50 minutes~\cite{azzopardi:IPAC, azzopardi:results}, as shown in Figure~\ref{fig:align_time_par}. Finally, this fully-automatic tool was also incorporated into the angular alignment implementation and successfully decreased the alignment time by 70\%, requiring no human intervention. Overall, the full-automation with the use of ML has proven to be more efficient and able to generate reproducible results. This is a major step in enhancing operational efficiency and will be used as the default software for starting the LHC in 2021.

\section{Optimisation of beam lifetime and losses}\label{sec: lifetime}
The LHC is a complex machine with numerous intertwined systems, each potentially impacting the dynamics and stability of the beams. As such, building a rigorous model of particle losses occurring in the LHC is a very daunting task, but it would offer valuable insight into the inner workings of the machine. This will help push its performance further, and also allow exploring ML techniques for potential use in the design and operation of a future FCC~\cite{Benedikt:2651300}.

The main goal behind this work is to develop a system capable of determining the optimal set of operational parameters so as to maximise the beams' intensity lifetimes given a specific machine configuration~\cite{Coyle:masterthesis}. This system could then assist in the setup of the LHC, and potentially help identify unknown correlations between different machine parameters. Eventually, the objective is to also compare the model built from experimental data with results from particle tracking simulations.

The approach we took to develop the system is to make use of the swaths of LHC data acquired through the several instrumentation systems in order to build a data-driven surrogate model of the beams' lifetime. This could then be coupled with an optimisation algorithm to determine the optimal operational parameters.

This problem was treated with a supervised-learning framework. The output of the model is the beam lifetime and the inputs are the operational knobs of the machine, i.e.\ the  tunes, chromaticities, and magnet currents. The data cover an entire operational year, but to simplify the input/output relationship, the data are taken from a small section of the complete machine cycle, corresponding to the end of the injection energy plateau just before launching the ramp. This will need to be extended in the future.

\begin{figure}
    \centering
    \includegraphics[width=0.8\linewidth,clip=]{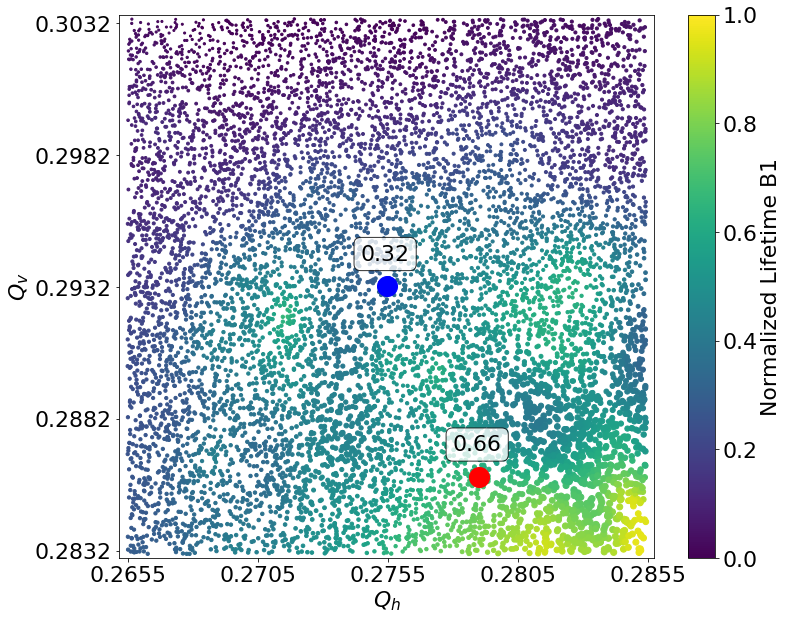}
    \caption{Normalised beam lifetime as a function of the LHC tune working point as measured for Beam~1. Blue dot: Nominal working point. Red dot: Lifetime-optimised working point as determined by the model. The model prediction is close, but not exactly equal, to the measured maximum lifetime. The absolute minimum and maximum value of the beam lifetime is 1.05~h and 32.7~h, respectively.}
    \label{fig:tune_diag_opti}
\end{figure}

Several SL models were trained and compared. The best performance was achieved with a Gradient Boosted Decision Tree model~\cite{NIPS2017_6907}. Once a surrogate model is trained, it can be paired with a variety of optimisers. In our case, an off-the-shelf simplex optimiser~\cite{NelderMead} was used to extract the optimal machine configuration from the trained lifetime response. We observed, however, that the distribution of the input data was far from ideal. This is to be expected, as LHC operation relies on reproducing strictly the same parameter set in every cycle to avoid uncontrolled beam losses with very high stored energy due to accidental `exploration' of the beam parameter space. By consequence, the surrogate model trained on the available data represents operational machine configurations well, but has a rather limited predictive power for non-operational machine setups. The parameter space was hence explored further with the help of a dedicated MD session in which multiple random tune scans were performed over varying machine configurations~\cite{Coyle:MD}. The data collected during this study are used to benchmark and supplement the current model. A number of beam instabilities that increased the beams' emittances, thus reducing the machine's performance, had been observed. Such instabilities are currently not taken into account by the model which is yet a weakness of the setup. Nonetheless, ignoring this blind spot and restricting ourselves to the, albeit naive, lifetime optimisation problem, the model does agree with the lifetime optimal regions of the vertical $Q_{\rm v}$ vs horizontal $Q_{\rm h}$ tune diagrams, see Fig.~\ref{fig:tune_diag_opti}. The model proves to be capable of moving towards the optimal regions, achieving a lifetime improvement of a factor of two, but falls short of the maximum. This could be due to the fact that the observed maximum is quite far from the nominal working point, as such the model will have sampled relatively few configurations as eccentric in the training data. The fact that the nominal working point does not coincide with the highest lifetime is due to the small distance between the tunes at the optimum lifetime point. It is currently not possible to ensure at all times sufficiently low linear betatron coupling and tune control to avoid accidental beam losses. It is however planned to move closer to the optimal working point in the coming LHC run.
To treat this problem rigorously this naive model will need to be built upon to take into account the emergence of instabilities in a multi-objective optimisation framework. Note that this work is a first approach to using data-driven ML techniques to develop surrogate models of beam lifetimes at the LHC. Although the models developed are still affected by limitations, nonetheless they are able to generate optimal values that are close to those obtained by the experimental measurements.
\section{Detection of collective beam instabilities}\label{sec: instab}
Collective instabilities can lead to a severe deterioration of beam quality, in terms of reduced beam intensity and increased beam emittance, and consequently a reduction of the collider's luminosity. It is therefore crucial for the operation of the LHC to understand the conditions in which they appear in order to find appropriate mitigation measures. For that purpose the LHC is equipped with a few dedicated measurement devices. Here we focus on the transverse damper's observation box (ObsBox)~\cite{obsbox}.

This device handles data coming from various transverse beam position monitors and keeps a rolling buffer of the data. When triggered, the ObsBox writes the buffer to disk. This trigger can be done manually, but the analyses presented here focus on data coming from an automatic system, which should ideally trigger the data saving when an instability is detected. The data saved is of very high resolution and contains bunch-by-bunch and turn-by-turn transverse beam position information throughout the machine cycle covering all beam modes, and fill types. The automatic triggering system has so far accumulated around $4$~TB of data. The analysis presented here focuses on the horizontal motion of Beam~1, but it can be trivially extended to the other beam and planes. This data set was acquired between $5^{th}$ of September 2017 and $3^{rd}$ of December 2018 and contains a total of 36196 triggers. Unfortunately, the vast majority of it does not actually contain any instabilities, which makes analysing the large amount of data a daunting task in itself. Currently the experts must painstakingly drudge through the data looking for instabilities in the hay stack.

In order to improve this search, a procedure capable of identifying unusual beam oscillation patterns occurring in the data and clustering the various kinds of signals together has been developed. Such a  clustering would drastically speed up the data analysis as the experts would be able to classify clusters of triggers at once, instead of sifting through the data to classify the individual triggers of the ObsBox.

\begin{figure}[htp]
    \centering
    \includegraphics[width=0.7\textwidth]{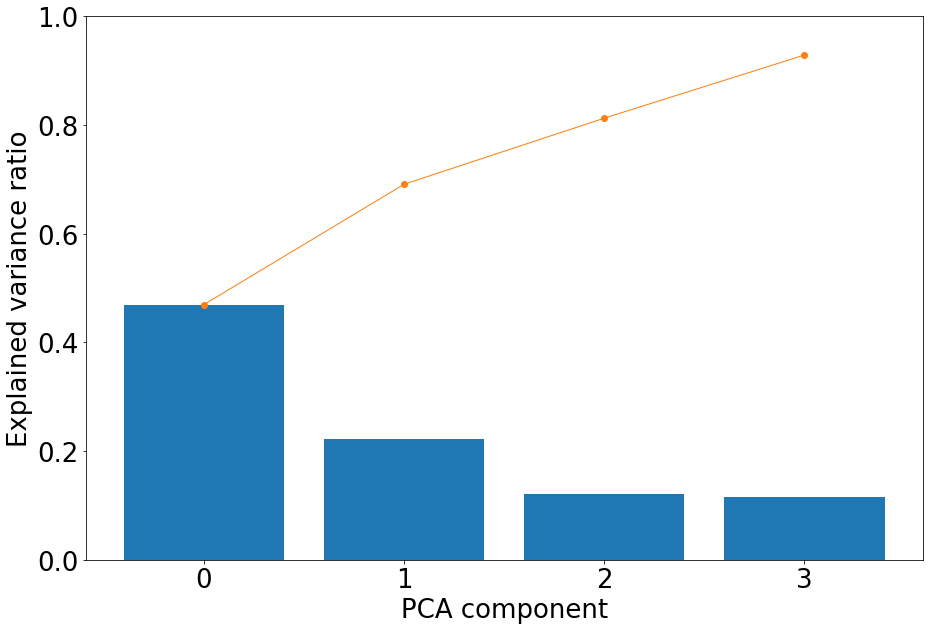}
    \caption{Principal Component Analysis of ObsBox automatic triggering data. Histogram: explained variance of each PCA component. Curve: cumulative explained variance.}
    \label{fig:PCA}
\end{figure}

Before any clustering can be performed, it is paramount to remove the false triggers from the data. This filtering step is crucial, as running computationally-expensive algorithms on the vast amount of falsely triggered data is unfeasible. This step can be seen as an anomaly-detection problem with the nominal samples being the false triggers and the anomalies the collective instabilities. The filtering is performed by extracting from data statistical features along with Fourier transform coefficients, Continuous Wavelet Transform coefficients, and the Complexity-Invariant Distance~\cite{CID} complexity estimate for a total of 29 features. The complexity estimate quantifies the complexity of a time series by stretching it out and computing the length of the flattened line. It is worth stressing that such a definition of complexity figure of merit is by no means the only one possible, but it has the advantage of being easily implementable and relatively cheap in terms of computing power needed. On this data set of extracted features, Principal Component Analysis (PCA)~\cite{wold1987principal} is performed, to obtain a projection to a lower-dimensional space by constructing new components which are linear combinations of the initial features. The dimension of this PCA space can be as low as $4$, while explaining 93\% of the variance of the extracted features, as shown in Fig.~\ref{fig:PCA}.

\begin{figure}[htp]
\centering
\subfigure[\label{fig:IF_inlier} Predicted inlier by Isolation Forest.]{
\includegraphics[width=0.70\textwidth, trim={0cm 0cm 0cm 1.1cm}, clip=]{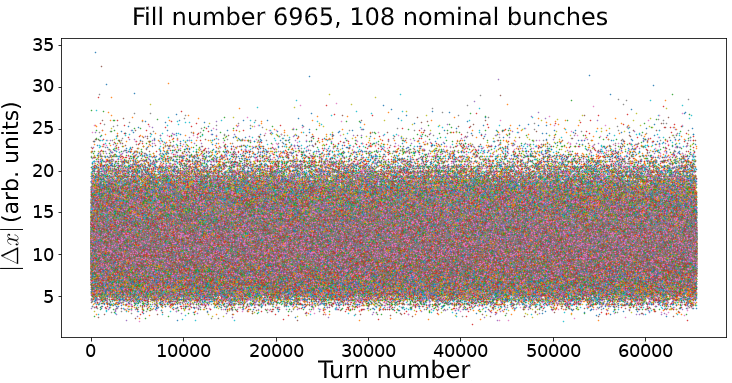}}
\subfigure[\label{fig:IF_outlier} Predicted outlier by Isolation Forest.]{
\includegraphics[width=0.70\textwidth, trim={0cm 0cm 0cm 1.1cm}, clip=]{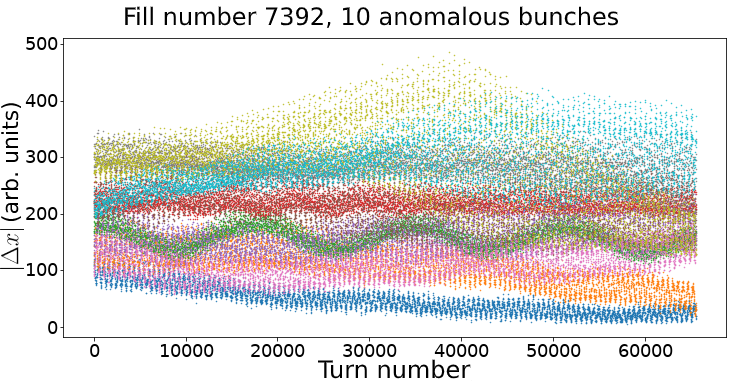}}
\caption{\label{fig:IF_preds} Examples of signals for 108 nominal bunches from fill 6595 (a) and 10 anomalous ones from fill 7392 (b), as predicted by the IF. $|\Delta x|$ is the absolute change in horizontal beam amplitude.}
\end{figure}

We then apply an off-the-shelf Isolation Forest~\cite{Liu:2008:IF} algorithm to such PCA space to isolate the anomalous samples. IF works by iteratively splitting the space using randomly placed hyperplanes, with the intuition that anomalous points take fewer iterations to isolate than nominal points, and based on this it is able to distinguish between the two. The quality of the IF's predictions is hard to evaluate quantitatively as there are no easily-accessible labels for these data. However, there is a small list of manually-classified instabilities~\cite{OBtable},  which was used to qualitatively tune and evaluate the accuracy of the IF. An example of a nominal and an anomalous signal, as predicted by the IF, is shown in Fig.~\ref{fig:IF_preds}. The nominal signals are from fill 6595, on $23^{rd}$ July 2018 at 22:07:21, and represent the situation during the injection beam process. On the other hand, the anomalous signals occurred in fill 7392, on $30^{th}$ October 2018 at 20:21:10, and represent the situation at top-energy, prior to bringing the beams in collision.

\begin{figure}[htp]
    \centering
        \includegraphics[trim= 5mm 5mm 5mm 5mm,width=0.57\textwidth,clip=]{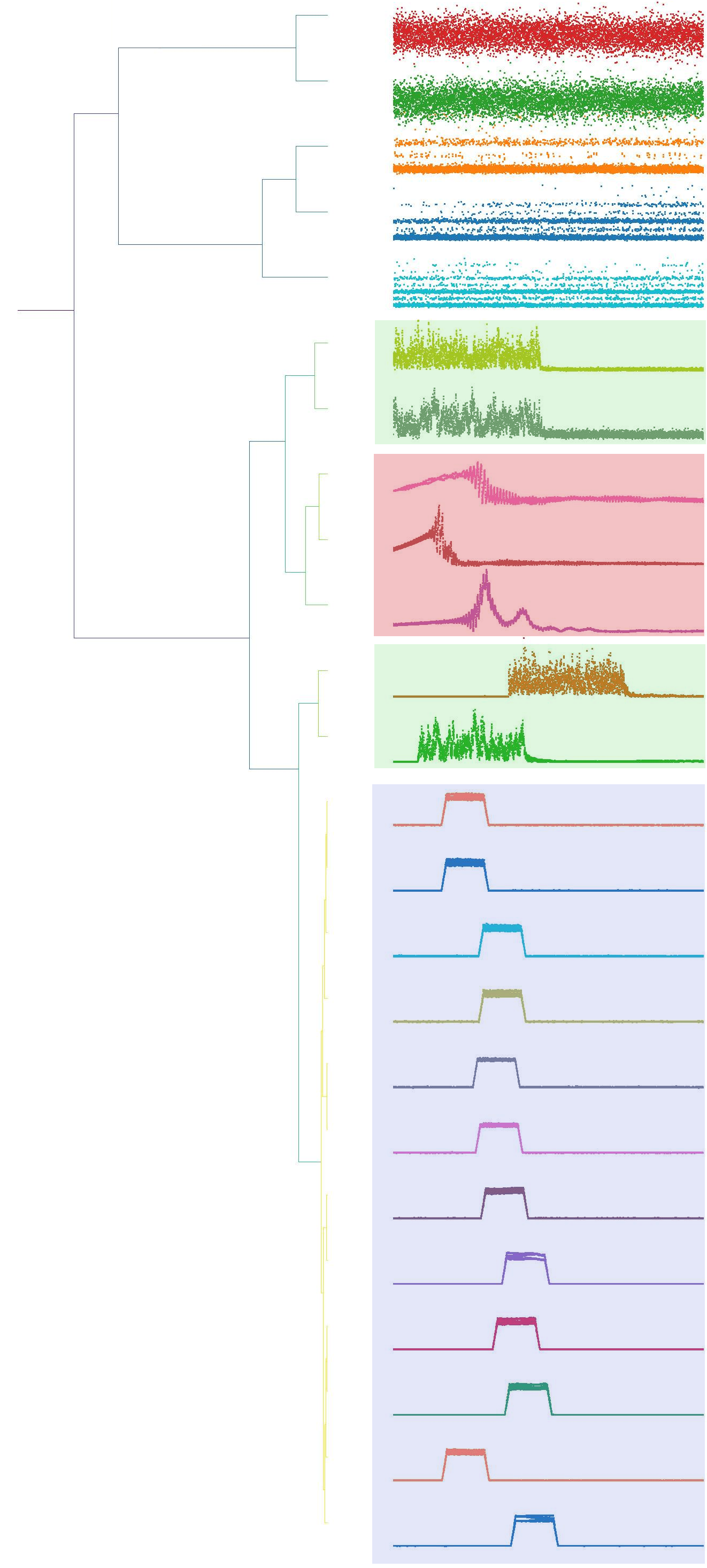}
    \caption{Hierarchical Clustering. Left: clustering dendrogram. Right: clustered signals. Highlighted are instabilities (red background area), beam offsets caused by the AC dipole (blue background area), a split cluster featuring transverse excitation (green background area), and an unknown behaviour (white background area). The smaller the dendrogram's branches the more similar two signals are.}
    \label{fig:clusters}
\end{figure}

Therefore, by using the trained IF we are able to filter out the false triggers from the ObsBox data. After retaining only the predicted anomalous samples from the entire set of ObsBox data, more computationally-expensive algorithms can be run for further analyses. In order to simplify the clustering, each bunch is considered as independent. Although with this assumption the model is not able to distinguish between single- and coupled-bunch instabilities, it greatly simplifies the modelling as we are left with a clustering problem for an univariate time series. The clustering of this type of time series requires determining an adequate distance metric, which should quantify how similar two time series are. A well-regarded metric for time-series similarity is the Dynamic Time Warping (DTW)~\cite{dtw_orig, dtw_review}, because, as shown in~\cite{DTW_compare}, DTW and its variants outperform other time-series similarity metrics and it is able to determine the similarity of two time series while remaining invariant to local warping. Using DTW it is possible to create a distance matrix describing the similarity of the evolution of each bunch with respect to all other bunches. This distance matrix is then fed to a Hierarchical Clustering Algorithm~\cite{hierar_clustering}, which iteratively identifies and links clusters to form a dendrogram. A random subset of such dendrogram, obtained from the whole data set, along with the corresponding signals, is shown in Fig.~\ref{fig:clusters}.

Signals with similar behaviour do, for the most part, get clustered together. However, it is clear that these data contain much more than simply transverse excitation due to collective instabilities. Indeed, there are other phenomena visible here. Notably, the step-function-like signals highlighted in blue background are caused by the AC dipole, installed in the LHC to measure the optical functions of the magnetic lattice~\cite{Serrano:1263248}. The AC dipole induces large transverse offsets of the beam that are clearly observed and clustered correctly. In the subset of plotted signals we do observe some well-clustered instabilities, highlighted in red background. It is also possible to observe some patterns that appear in different clusters, while they should be put together, which are highlighted in green background. This shows a limitation of the DTW implementations used, which does not allow for partial matches. Other implementations can relax the end-point conditions, which could be used if deemed necessary. The remaining signals in this random subset are clustered correctly, however, the cause of their behaviour remains unknown, and further investigations must be performed to explain these patterns. It is clear from this analysis that the ObsBox's triggering system can be improved. In fact using a data-driven model, similar to those presented, to control the triggering in a more intelligent manner would drastically reduce the number of false triggers.

\section{Heating detection from pressure readings}\label{sec: heating}
The charged particle beams stored in a high-energy, high-intensity accelerator such as the LHC may induce heating of surrounding equipment. The main sources of heating are: electron cloud~\cite{rumolo2003electron}, particles lost on the beam surroundings~\cite{bruce2014simulations}, synchrotron radiation~\cite{sokolov1966synchrotron}, and beam-induced RF heating due to impedance~\cite{zannini2014power}, which has been one of the limitations to reaching nominal performance of the machine during the LHC Run~1~\cite{salvant2013update}.

In principle, the beam-induced heating can be directly monitored by means of temperature probes, such as PT100~\cite{vega2017thermal} devices in the LHC ring or optical fibres in the CMS~\cite{saccomanno2012long} detector. However, large fractions of the LHC ring are left without temperature monitoring. Temperature increase in a high-vacuum environment may lead to outgassing~\cite{grobner1999dynamic}, which can be observed as pressure increase in vacuum gauges and this was the case, e.g. for the injection protection device (TDI), during the 2012 LHC run~\cite{salvant2013update}. However, this was not an isolated case. Indeed, abnormal outgassing levels were observed in the Beam Gas Ionization (BGI) profile monitor during Run~1 and the installation of temperature probes confirmed during Run~2 that the outgassing was in fact linked with heating effects. Similarly, some vacuum modules implementing RF fingers featured high-level of pressure spikes, which turned out to be due to RF heating and could be solved by an improved design of the modules. Overall, the behaviours observed during LHC operations hint clearly to the need of an efficient tool for an early detection of heating effects to avoid serious damages to the LHC hardware.

Vacuum monitoring is much denser and systematic than temperature monitoring, with more than 1200 vacuum gauges distributed all over the LHC circumference~\cite{jimenez2009lhc}. Nevertheless, analysing the pattern of the readings from these vacuum gauges one by one after each LHC fill in order to detect abnormal behaviour, represents a tedious and time-consuming task. Not to mention that a robust technique to translate the vacuum readings into equivalent values of temperature is not completely trivial. Hence, automatising this pattern-classification process is expected to result in a significant gain of time for the physics run and manpower.

\begin{figure}[!ht]
    \centering
    \includegraphics[width=1.0\linewidth,clip=]{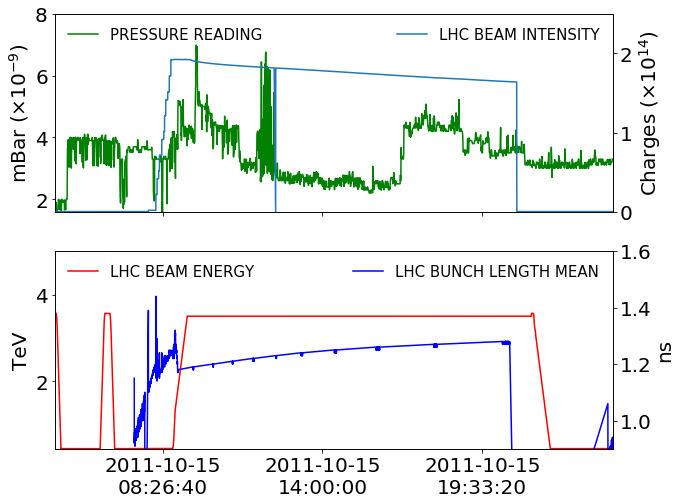}
    \caption{Example of the information used to estimate beam heating effects from a pressure reading, in this case from a vacuum gauge located in Sector~4-5, close to the stand-alone magnets D4 and Q5, on the Beam~2 channel of the LHC, during a physics fill of the 2015 run. The beam intensity and the pressure reading are shown in the top plot, whereas the beam energy and the average bunch length are shown in the bottom plot. The sudden pressure increase during the LHC cycle can be an indication that beam-induced heating is taking place in the surroundings of the vacuum gauge.}
    \label{fig:heating_example}
\end{figure}

The applicability of ML to this task has been investigated by building an automatic classification algorithm for pressure readings produced by the vacuum gauges in order to detect heating patterns. Heating in a pressure-reading pattern can be observed as an anomalous pressure increase as seen in Fig~\ref{fig:heating_example}. The pressure evolution of a vacuum gauge located in Sector~4-5, close to the stand-alone magnets D4 and Q5, on the Beam~2 channel is shown as a function of time during a typical LHC cycle. The evolution of beam intensity, energy, and the average (over the bunches) bunch length is also added. The pressure reading features sudden changes that are not related with bunch length variations, thus hinting to outgassing induced by a temperature increase. It is clear that in this approach the underlying physical phenomenon triggering the temperature rise is disregarded and only its consequences are analysed.  

Since the goal is to reduce the time needed in finding the abnormal gauges, the classifier aims to select a subset of all the gauges producing data in which all the abnormal ones are present. Statistically speaking, this means that the classifier should reach a high $recall$ score~\cite{goutte2005probabilistic}, which is defined as the fraction of true positives detected over the total amount of positive cases. In order to apply SL techniques, more than 700 readings have been labelled with expert supervision creating a data set with 700 time series and of 3000 steps in time each, where each time series is a vacuum gauge reading. Even if the data set is made of time series the goal here is to classify them and not to predict future values as it is usually done with common time-series problems.

To reduce the dimensionality of the data set, a PCA~\cite{wold1987principal} has been performed leading to retain only 12 features. These 12 features do not have any physical meaning, but they explain the $99.9\%$ of the variance of the full data set. In this way, the dimensionality reduction does not lead to a significant information loss. On the resulting data set with only 12 features, firstly a K-Nearest Neighbour Classifier (KNN)~\cite{dudani1976distance} and then Multi-Layer Perceptron (MLP)~\cite{pal1992multilayer} have been trained. For the KNN algorithm, the best value of $k$ and the best algorithm to use between k-dimensional Tree (k-d Tree) and Ball Tree~\cite{otair2013approximate} and a boolean value that indicates whether the data set has been scaled to have zero mean and unit variance have been tuned using Grid Search~\cite{lavalle2004relationship}. For the MLP algorithm, a sigmoid activation function (also called logistic function) has been applied at the output layer to perform a classification, while a rectified linear unit activation function, defined as the positive part of its argument, i.e. $max \{ 0, x \}$ where $x$ is the input value, has been used for the hidden layers. Cross-entropy has been used as loss function being a classification task. Randomised Search~\cite{bergstra2012random} has been applied to identify the optimal number of layers and neurons in order to explore a wide range of input values. 
The goal of the tuning of hyper-parameters was to increase the $recall$ score.

To evaluate the performance of both the KNN and the MLP classifiers, a 4-fold cross-validation technique~\cite{krogh1995neural} has been applied while training each model. Stratified splitting~\cite{neyman1992two} has been used for the 4-fold technique in which the folds are made by preserving the percentage of samples for each class.

\begin{figure}[!ht]
    \centering
    \includegraphics[width=1.0\linewidth,clip=]{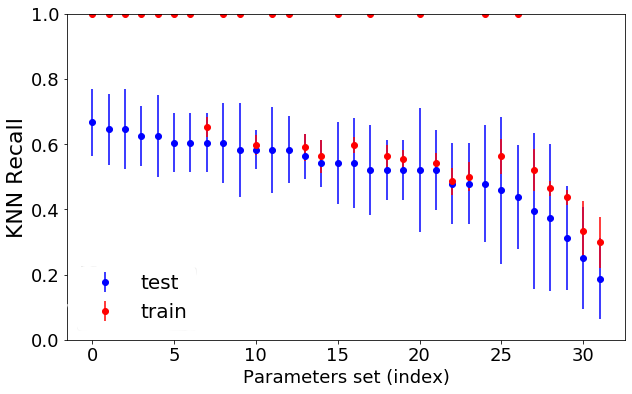}
    \includegraphics[width=1.0\linewidth,clip=]{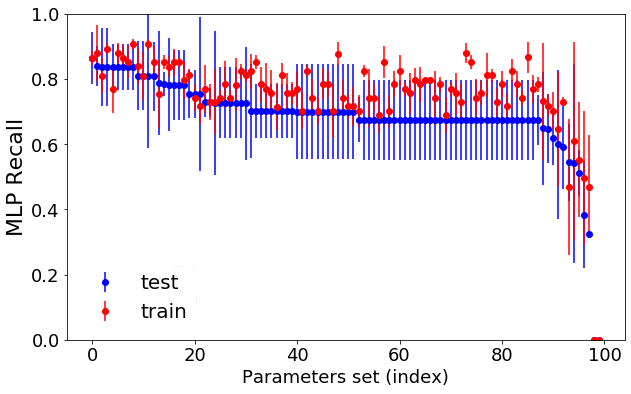}
    \caption{$Recall$ scores of the KNN classifier (top) and of the MLP (bottom). The red dots at $recall=1$ in the KNN classifier are parameters set for which the algorithm is overfitting the training data.}
    \label{fig:KNN_MLP_Recall}
\end{figure}

The results of the parameter set scan maximising the $recall$ for the KNN are shown in Fig.~\ref{fig:KNN_MLP_Recall} (upper). Note the red dots corresponding to $recall=1$ in the KNN classifier are parameter sets for which the algorithm is overfitting the training data. In general a model is overfitting the training set if it behaves extremely well on them, but then behaves poorly on the test data and $recall=1$ for the training data is a perfect example of overfitting.

The best parameter set is found to be that at index 7 in the plot, i.e the first set of parameters that is not overfitting the training set, for which $recall=0.60 \pm 0.09$. The parameter set for the the KNN contains the value $k$ of the algorithm, the algorithm used, and a boolean value that indicates whether the data set has been scaled to have zero mean and unit variance. For the best parameter set $k=5$, the algorithms is k-d Tree, and the data set has been scaled. Note that KNN has $precision = 0.97 \pm 0.04$ meaning that within the vacuum gauges classified as abnormal, 97\% are actually abnormal.

The MLP scan over the network parameters is represented in Fig.~\ref{fig:KNN_MLP_Recall} (bottom). This result is achieved for the parameter set index 0, corresponding to a network made of 2 hidden layers with 176 neurons per layer. The $recall$ score for the neural network is $0.86 \pm 0.10$, which represents a $35\%$ improvement with respect to the KNN case. The $precision$ score in this case is $0.58 \pm 0.04$, which is lower than the KNN $score$. However, MLP is to be preferred over KNN, since $recall$ is the main figure-of-merit and the increased computational burden is not an issue, given the relatively small data set used for training.

The implementation of a very simple Neural Network shows already promising $recall$ scores, which motivates testing more refined ML techniques on this task. Convolutional Neural Network~\cite{krizhevsky2012imagenet} and ensemble methods are currently being investigated to push further the performance of the classifier.
\section{Numerical simulations of dynamic aperture}\label{sec: DA}
\begin{figure}[!ht]
    \centering
    \includegraphics[width=42mm]{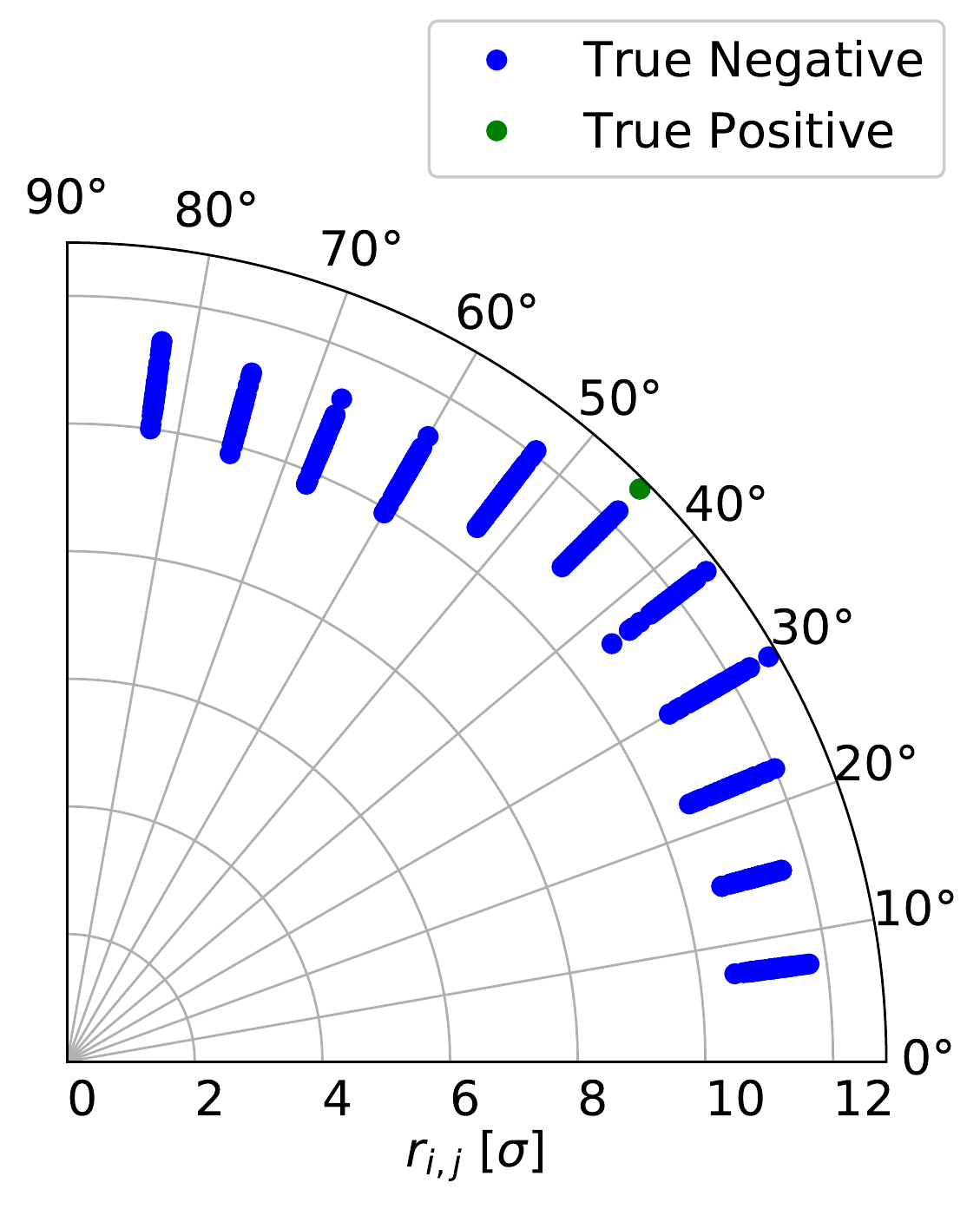}
        \hspace{1.5truecm}
    \includegraphics[width=42mm]{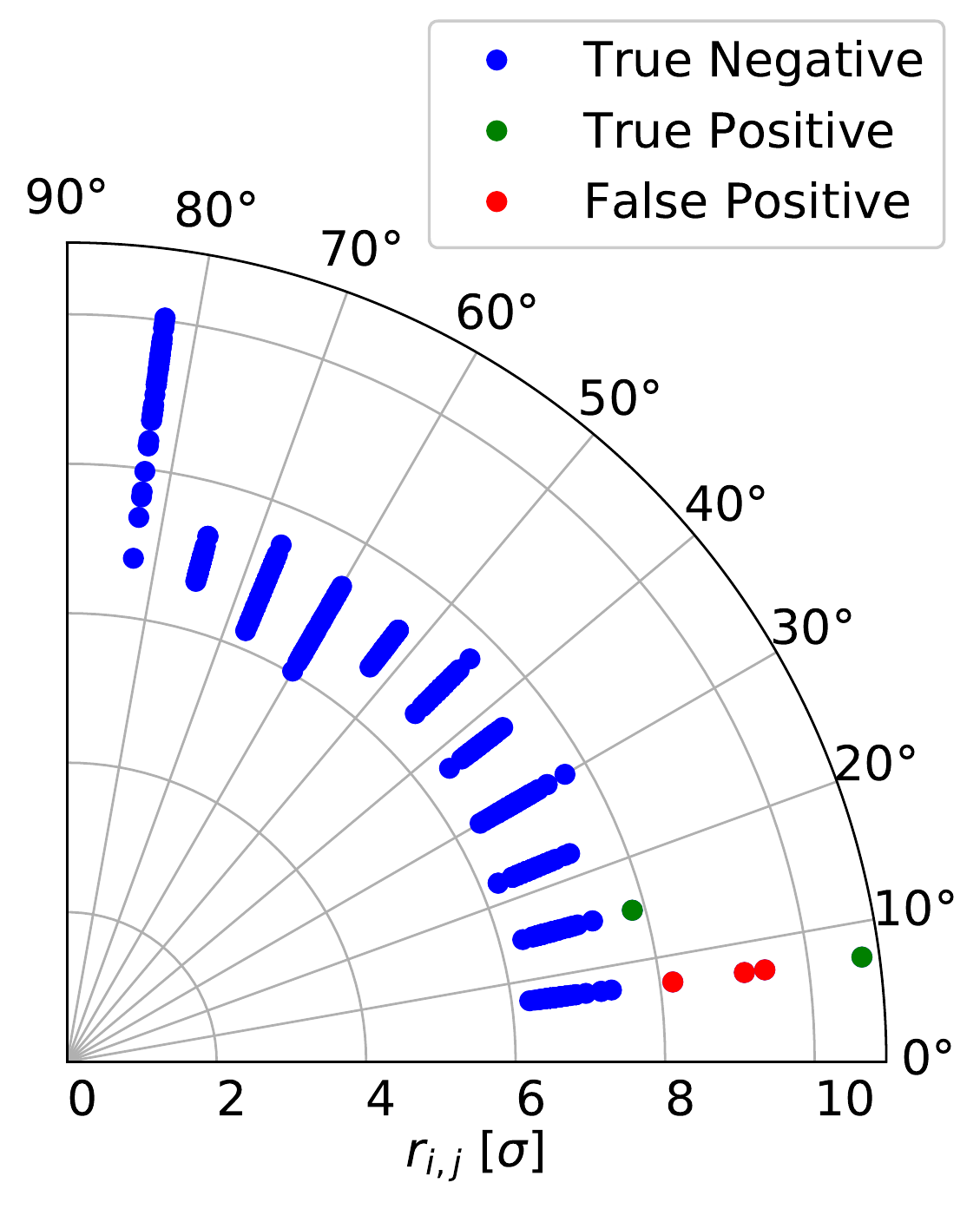}
    \caption{DA simulations for two LHC configurations. The initial 4D co-ordinates are of the type $(x,0,y,0)$ and a polar scan is performed on $(x,y)$. The various markers represent the results of the sixty seeds used. Left: Example of a DA computation where an outlier is correctly flagged (in green). Right: Examples of false positives (in red). The FP cases, are less worrisome as they refer to the determination of the maximum stable amplitude, which does not affect ${\rm DA_{min}}$.}
    \label{fig:DAoutliers}
\end{figure}

One of the most relevant and useful concepts in the study of non-linear beam dynamics is that of Dynamic Aperture (DA), which represents the radius of the smallest sphere inscribed in the connected volume in phase space in which motion is bounded over a given time interval~\cite{DAdef}. It can be estimated from tracking simulations, where a given set of initial conditions, uniformly distributed in polar co-ordinates in normalised physical space, are probed for bounded motion. All this is repeated for a number of different realisations of the magnetic field errors (the so-called seeds) for a given accelerator model, according to the following formula
\begin{equation}
{\rm DA_{ave}} = \frac{1}{N_{\rm seed}} \sum\limits_{i=1}^{N_{\rm seed}}  \int\! d\theta \; r_i(\theta) \, ,
\end{equation}
where $r_i(\theta)$ represents the last stable amplitude for seed $i$ in the direction given by the angle $\theta$. This definition is typically used for a refined understanding of the features ruling the DA (see, e.g.~\cite{Bazzani:2019csk}). However, for design studies, where conservative estimates are more appropriate, the DA is evaluated as
\begin{equation}
{\rm DA_{min}}=\min_{i,j} r_{i,j} \quad 1 \leq i \leq N_{\rm seed} \, , \, 1 \leq j \leq N_{\rm angle} \, ,
\label{eq:DA}
\end{equation}
where $r_{i,j}$ represents the last stable amplitude for the $i$th seed and $j$th angle. Such a definition might be strongly affected by outliers, which is the reason for our attempts to provide automatic tools to deal with outlier recognition. An example of DA plots is given in Fig.~\ref{fig:DAoutliers}, where results of DA computations for two LHC configurations are given for sixty seeds, eleven angles, and $10^5$ simulated turns. The left plot refers to the optics version 1.3 for the High-Luminosity LHC at top energy, with $\beta^*=15$~cm, $Q^\prime=15$ and strong powering of the Landau octupoles, but without beam-beam effects. The right plot refers to the optics configuration of the LHC during the 2016 proton run at injection, $Q^\prime=8$ and strong Landau octupoles to fight electron-cloud effects. 

It is not uncommon that for a given angle the stable amplitude differs considerably from seed to seed, thus generating a distribution of stable amplitudes over seeds with outliers, which can strongly affect  ${\rm DA_{min}}$. Outliers may arise  because the distribution of non-linear magnetic errors excites particular resonances in a way that is highly seed-dependent. It is clear that outliers possibly represent unlikely configurations that might be removed from the analysis of the numerical data for the computation of ${\rm DA_{min}}$. ML techniques have been used in large-scale DA simulations to flag certain results as outliers, which can then be dealt with accordingly.

One has to make sure to distinguish a set of outliers from a justifiable split of a set of points into a certain number of clusters. For this reason, the outlier detection is done in several steps. First, for each angle the $r_{i,j}$ values for that angle and for the different seeds are rescaled between the minimum and maximum values. We then investigated two types of ML approaches in order to automatically detect outliers. In the SL approach, we treat the goal of outlier detection as a classification problem, and train a SVM to distinguish between normal and abnormal points. Following a hyperparameter search, we identified the Radial Basis Function (RBF) kernel~\cite{RBF} with a penalty factor $C$ of unity as the best hyperparameters for the SVM model. 

\begin{figure}[!ht]
    \centering
    \includegraphics[width=84mm]{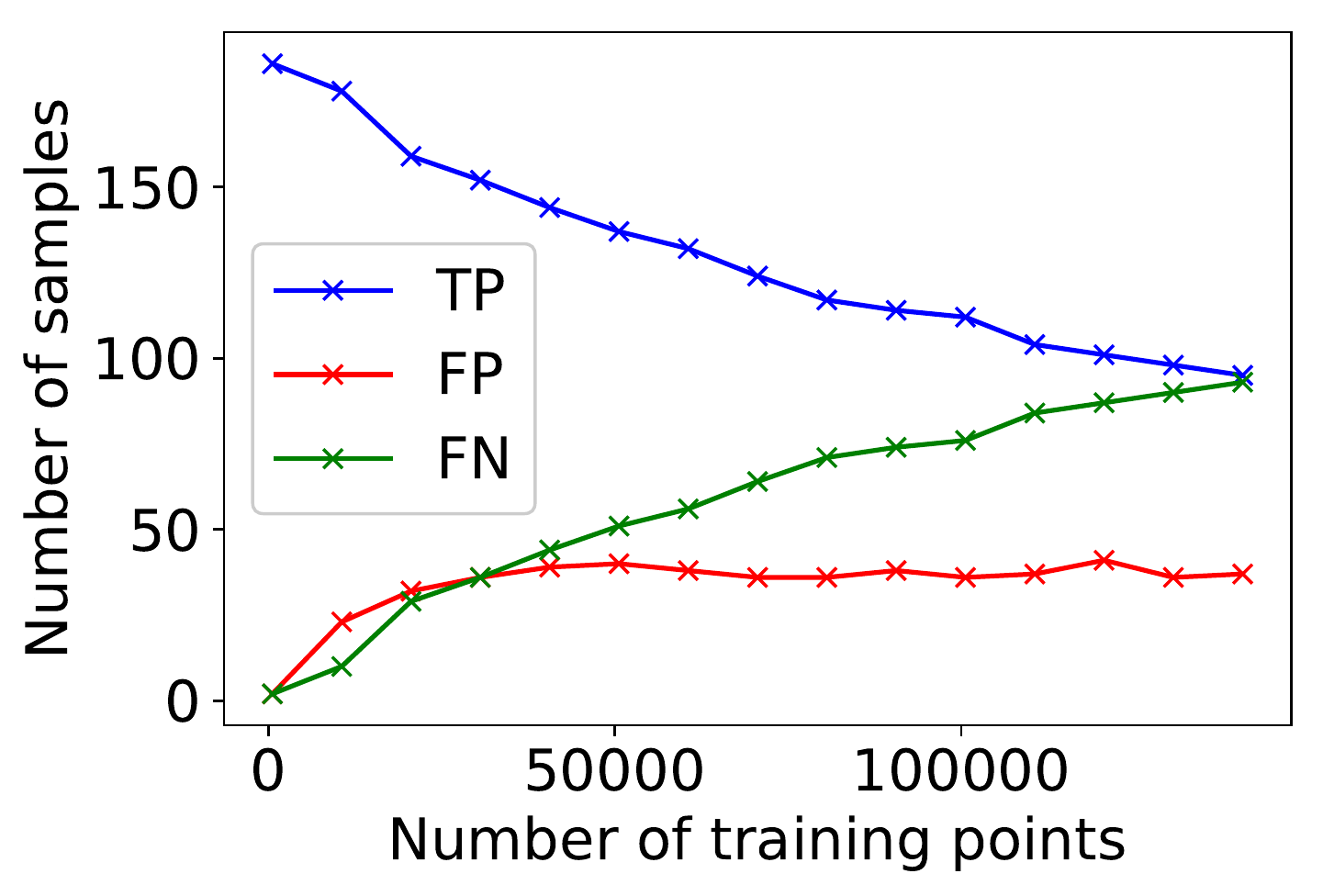}
    \includegraphics[width=84mm]{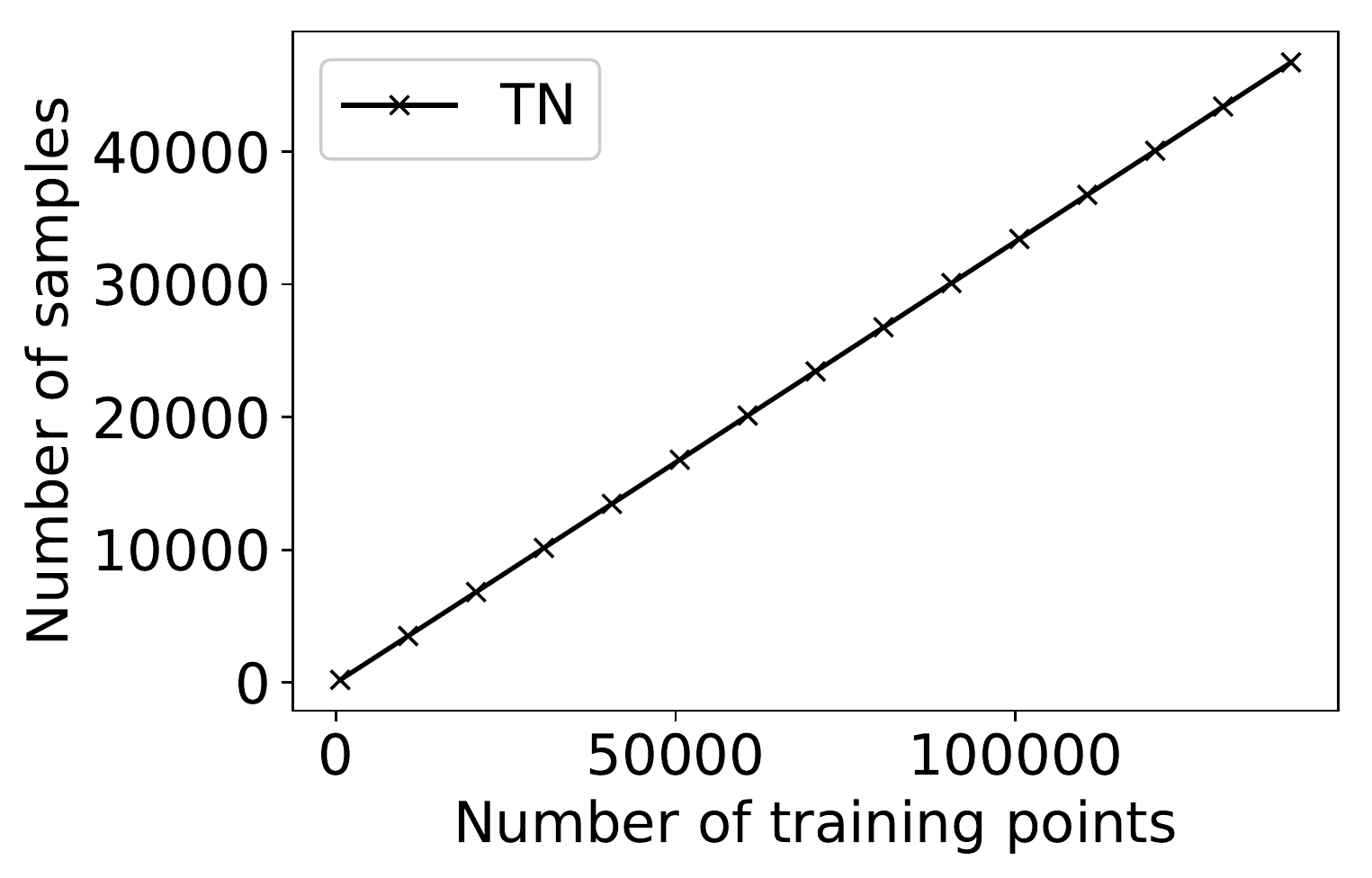}
    \caption{Learning curves for the SVM training, showing the TP, FP and FN (top) and the TN (bottom).}
    \label{fig:learning-curve}
\end{figure}

It is useful to observe the performance of the model as a function of the number of training points. This is known as a learning curve, and is shown in Fig.~\ref{fig:learning-curve}. Each point in the curves represents the number of True Positives (TP), False Positives (FP), True Negatives (TN) and False Negatives (FN) obtained on a test data set whose size corresponds to 25\% of the overall data set available, when the model is trained on all anomalous points plus a certain increasing number of normal points. A TP corresponds to a ground-truth anomalous point, which was correctly predicted to be anomalous. The results show that when the training data set is close to being balanced between anomalous and normal points, the number of TP is quite high, while the FP and FN are low. However, as the data set becomes more and more skewed towards normal points, the model achieves a lower performance. This is understandable given the assumption of balance in the SVM algorithm. 

We also investigated two UL approaches for detecting anomalies on an angle-by-angle basis. The first algorithm is the Density-Based Spatial Clustering of Applications with Noise (DBSCAN) method~\cite{Ester96adensity-based}. The DBSCAN is a density-based clustering non-parametric algorithm: it groups points that are closely packed together. The points which are not assigned to any cluster after applying the algorithm are automatically considered to be outliers. The second approach is the Local Outlier Factor (LOF)~\cite{LOF} algorithm. LOF quantifies the outlier strength of each point based on a concept of a local density. Locality is given by the $K$ nearest neighbours, whose distance is used to estimate the density. The comparison of the local density of an object to the local densities of its neighbours, allows regions of similar density to be identified. The points that have a substantially lower density than their neighbours are considered to be outliers. 

Following hyperparameter optimisation, the following is a list of the determined hyperparameters for each method:

\begin{itemize}
    \item DBSCAN: eps=1 (the maximum distance between two samples for one to be considered as in the neighbourhood of the other); min\_samples=3 (the number of samples, or total weight, in a neighbourhood for a point to be considered as a core point, including the point itself);
    \item LOF: n\_neighbors=58 (number of neighbours used to measure the local deviation of density of a given sample with respect to the same neighbours); contamination=0.001 (the expected proportion of outliers in the data set).
\end{itemize}

\begin{figure}[!ht]
    \centering
    \includegraphics[width=84mm]{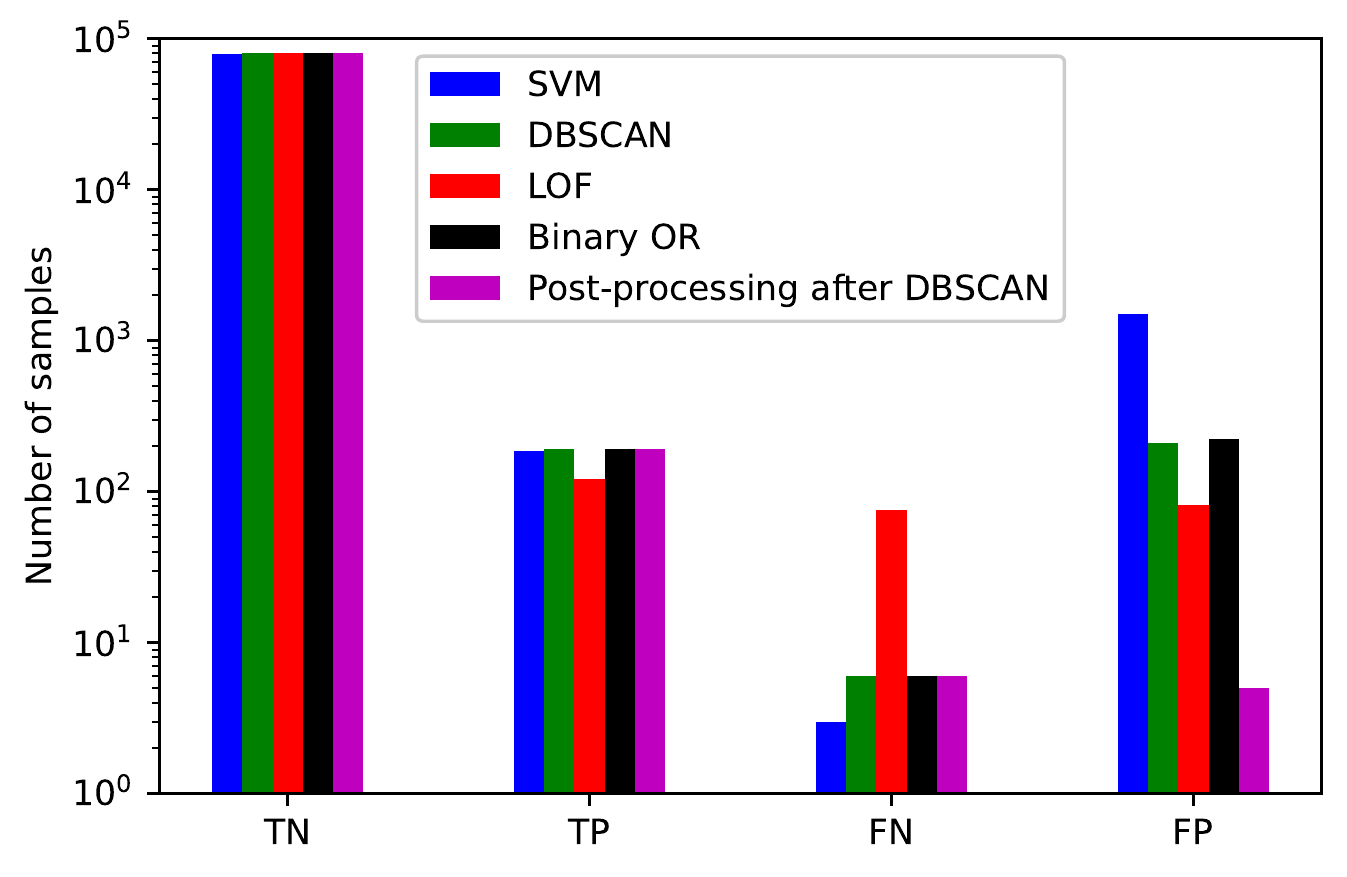}
    \caption{Results from anomaly detection using SVM, DBSCAN, LOF, a binary OR between DBSCAN and LOF and post-processing following DBSCAN methods. TP = True Positives (anomaly correctly detected), TN = True Negatives (normal point correctly detected), FP = False Positives, FN = False Negatives.}
    \label{fig:confusion-matrices}
\end{figure}

A comparison between the SVM, DBSCAN, and LOF algorithms is shown in Fig.~\ref{fig:confusion-matrices}. The labels predicted by the DBSCAN and LOF algorithms were also combined through a binary OR operation to produce a fourth set of labels. A further fifth set of labels is created by removing false positives using a statistical method (following an initial labelling by DBSCAN) to determine whether this would add to the robustness of the prediction. For a point, being flagged by DBSCAN as an outlier, to be considered a true outlier, we demand that it satisfies three additional criteria: the distance from the mean should be at least 3 standard deviations (where mean and standard deviation are calculated over the regular points, only); the distance to the nearest regular point should be at least $0.15$, in absolute units; the distance to the nearest regular point should be at least 34\% of the total spread of the regular points. This post-processing is performed in an iterative manner, starting at the minimum (maximum) point and working outwards (inwards),  recalculating the statistical variables of the regular points at every step. The values of these thresholds are chosen empirically as such to help catching FP points that arise from strongly-packed clusters.

The results show that the unsupervised methods perform an order of magnitude better than SVM in terms of false positives, however they are worse in terms of false negatives, especially when using LOF. The method of post-processing following DBSCAN clearly contributes to reducing the number of false positives, while maintaining the TP and FN rates.

\begin{figure}[!ht]
    \centering
    \includegraphics[width=84mm]{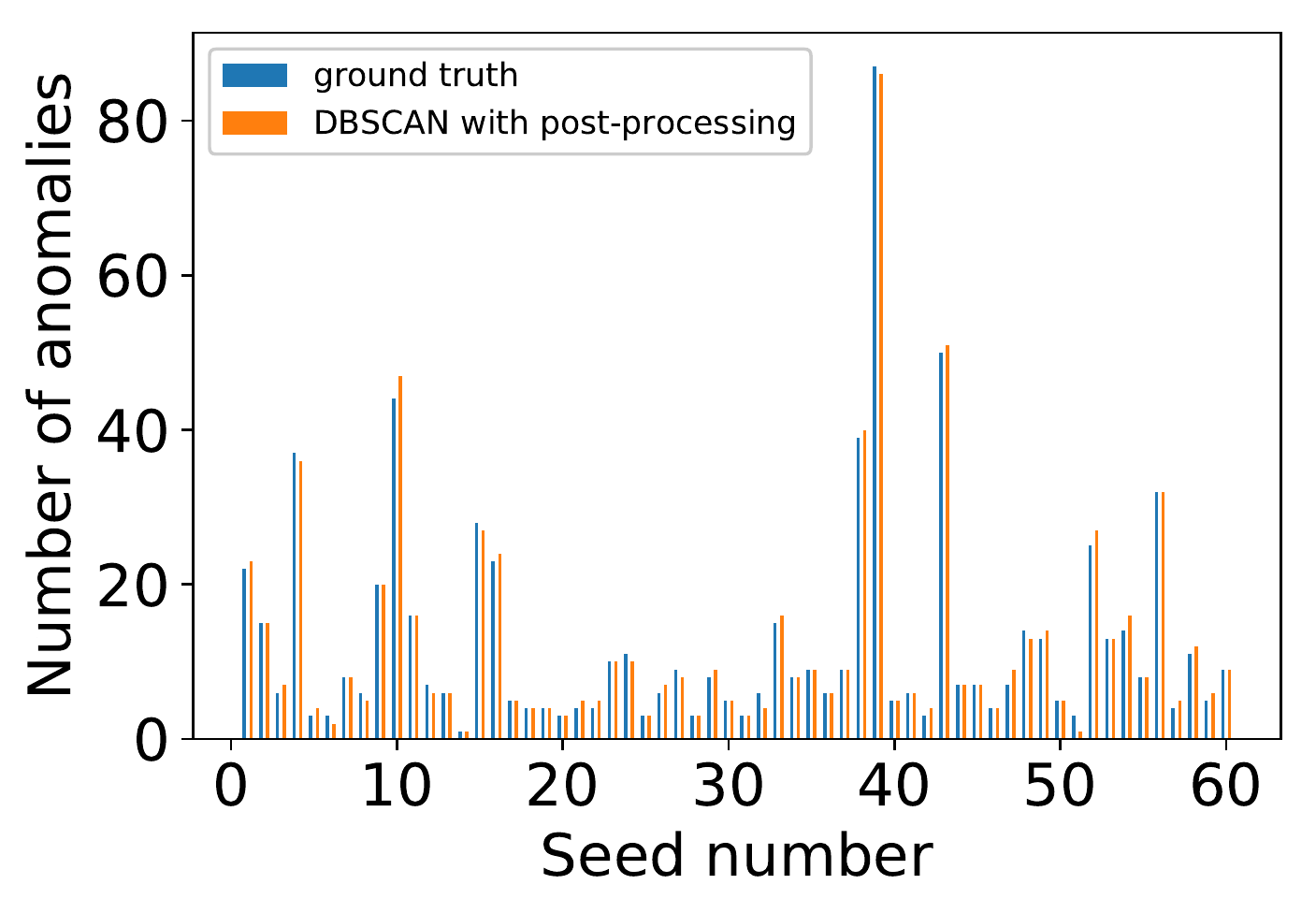}
    \includegraphics[width=84mm]{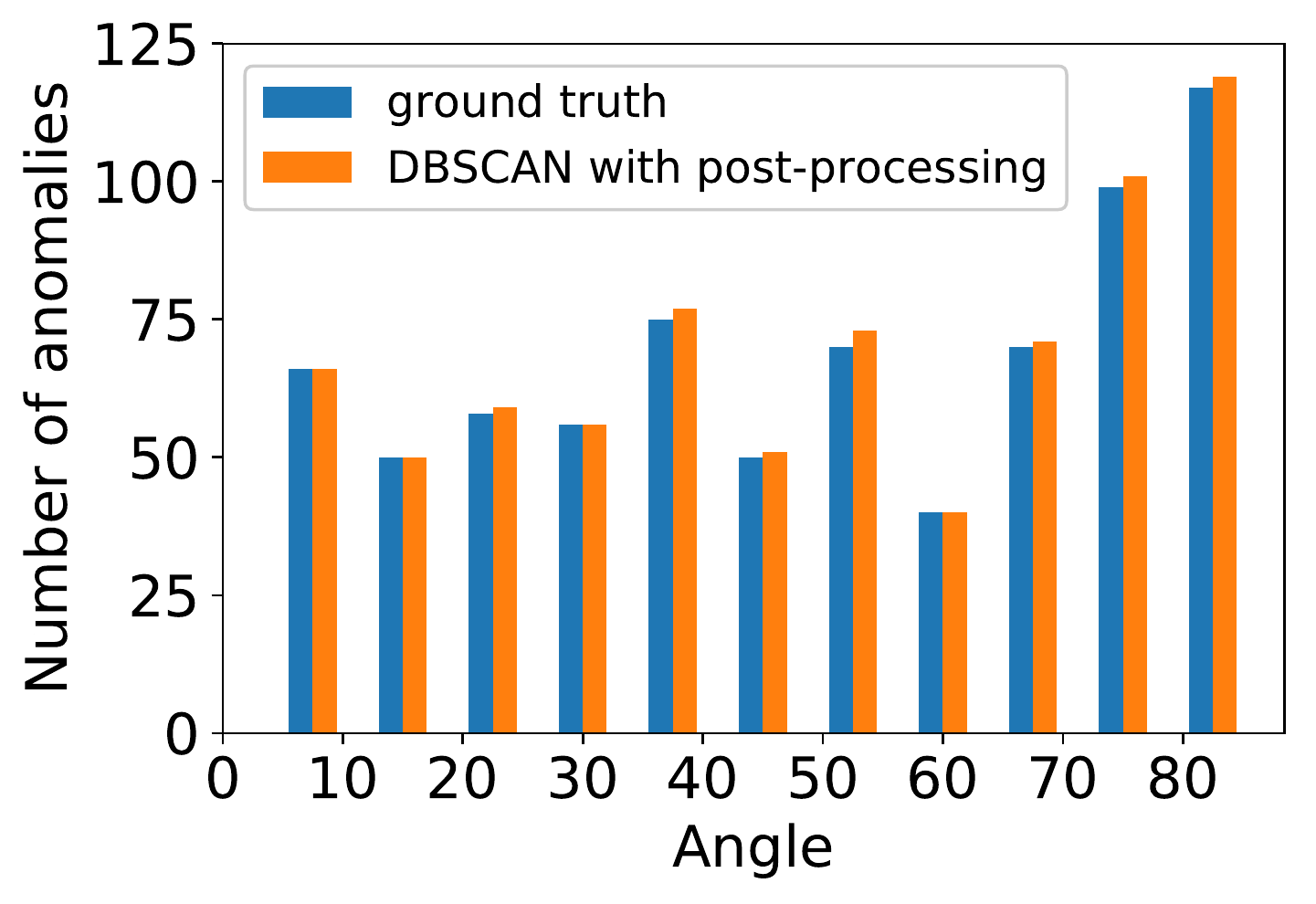}
    \caption{Visualizations of the anomalies by seed (top) and by angle (bottom), showing the similarity between the ground-truth and the result of the post-processing following DBSCAN.}
    \label{fig:anomalies-by-seed-and-by-angle}
\end{figure}

\begin{figure}[!ht]
    \centering
    \includegraphics[width=84mm]{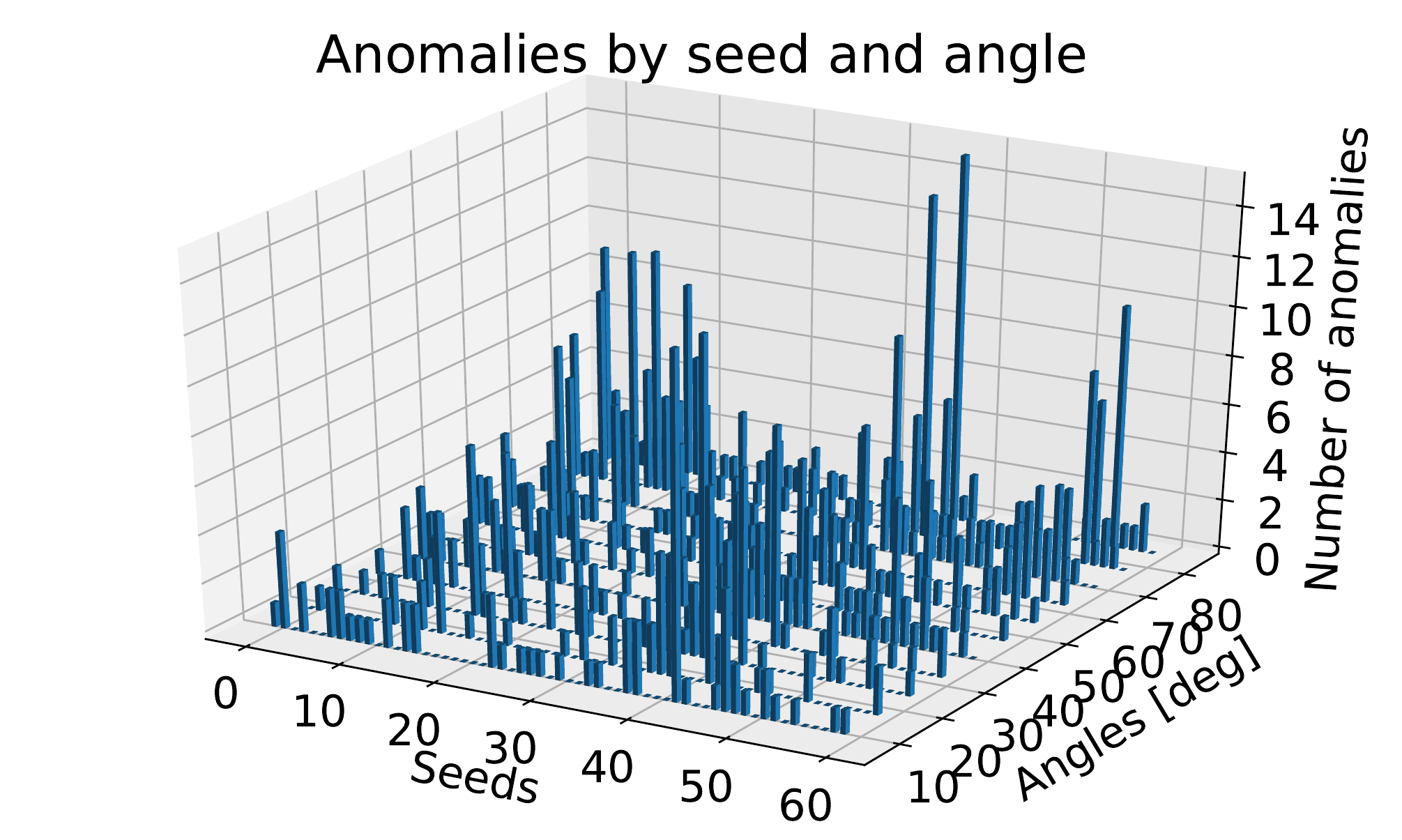}
    \includegraphics[width=84mm]{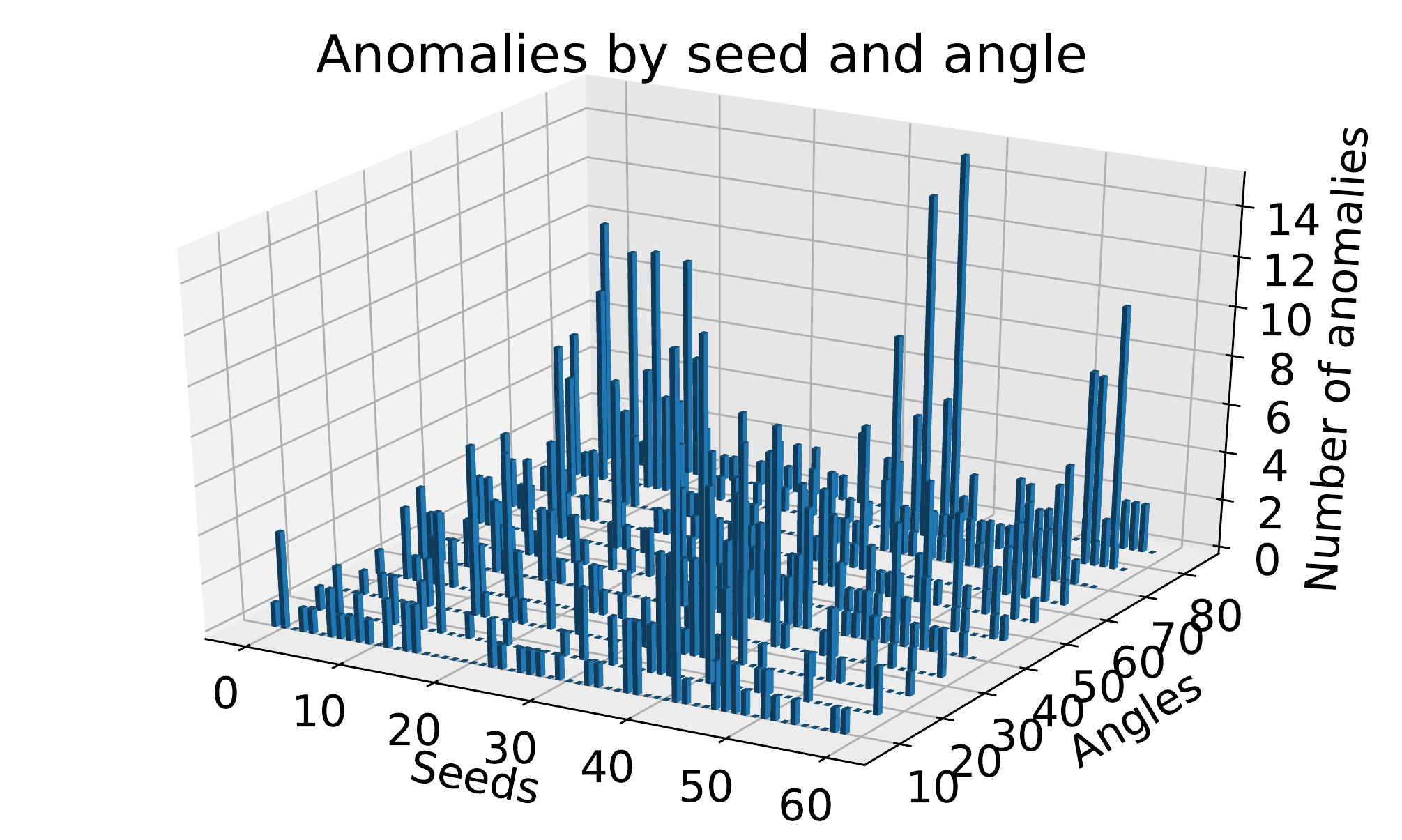}
    \caption{Visualizations of the anomalies by seed and angle for the ground-truth (top) and the result of the post-processing following DBSCAN (bottom).}
    \label{fig:anomalies-by-seed-and-angle}
\end{figure}

It is also very useful to investigate the dependence of the number of outliers on the angular distribution and on the seed number. A comparison between the ground-truth and predicted anomalies (using post-processing following DBSCAN) is shown in Fig.~\ref{fig:anomalies-by-seed-and-by-angle}, where it can be seen that the anomaly profiles for both seeds and angles are similar between the ground-truth and the predictions. It is worth noting the peculiar profile of outliers as a function of seed, featuring three clusters with very large number of outliers. As far as the anomalies distribution as a function of angle is concerned, there is a tendency towards larger number of outliers for angles close to $90^\circ$. As a last observation, the number of outliers at low amplitude equals that at large amplitude. A combined visualisation for seeds and angles is shown in Fig.~\ref{fig:anomalies-by-seed-and-angle}. 

This analysis potentially provides insight into the sensitivity of the underlying physics; investigations on this matter are still on-going.

Two examples of the classification obtained by means of the post-processing following DBSCAN are shown in Fig.~\ref{fig:DAoutliers}. 

It should be noted that the dynamics governing the DA can be very different as a function of the angle $\theta$, hence even when the neighbouring points are similar in amplitude, the spotted outlier might be a genuine one. It is clear that particular care needs to be taken in these cases before drawing any conclusions, and additional investigations might be advisable.
\section{Conclusions and outlook} \label{sec: conc}
In this paper, a selection of ML applications, based on algorithms of either Supervised Learning or Unsupervised Learning type, for  a variety of domains linked with beam dynamics aspects at the LHC have been presented and discussed in detail.

All this started from the quest for improving optics measurements and corrections, in particular by detecting faulty BPMs in harmonic analysis of turn-by-turn measurements to avoid the appearance of outliers in the computed optics functions. Data cleaning has also been successfully achieved by means of anomaly detection and clustering techniques. As far as optics corrections are concerned, already the basic Neural Network implementation produces very interesting and promising results. Note that a larger data set is being generated and more error sources and non-linearities are being added in order to create a more general model. Further improvements are foreseen by using an autoencoder network to improve the quality of betatron-phase measurements, which is the fundamental part of optics and corrections computations. Furthermore, autoencoders could be used to perform denoising of turn-by-turn data.

Excellent results have been obtained with the automated alignment of collimators, in which ML has been used to distinguish genuine beam-loss spikes from spurious events during the alignment process of the collimators' jaws. This approach has achieved a remarkable speeding up in setup time of the collimators systems, with an overall beneficial impact on LHC operation as the ML implementation has become the operational one. A continued analysis of the cross-talk effects for the losses of the two beams in the future will allow to perform more alignments in parallel and is being pursued actively. 

The ML model of the LHC beam lifetimes showed promising results once the parameter space was uncorrelated by a dedicated machine experiment in the LHC. While it manages to represent operational machine setups well, it lacks predictive power when considering non-standard configurations at present. Future work includes exploring alternative ML approaches, such as treating the data as time series, evaluating the model performance across several years of LHC operation, and supporting the experimental data with a surrogate model based on detailed beam loss simulations from particle tracking codes.

Unsupervised learning has been employed to detect and classify beam instabilities as well as other anomalies in the beam position data acquired by the LHC ObsBox system through an automatic triggering system. Promising results have been obtained by combining anomaly-detection techniques with a Hierarchical Clustering algorithm. Dynamic Time Warping was found to provide a robust distance metric to group similar-looking time series. The results demonstrate the power of these techniques when dealing with drastically unbalanced data sets, i.e. where only a small fraction of the data actually represents beam instabilities. To further improve the analysis, potentially important changes are already envisioned, such as using autoencoders in place of the feature extraction, Principal Component Analysis and Isolation Forest steps. Additionally, it will be explored whether the ML approach can help improving the automatic ObsBox triggering with the goal to reduce the number of false triggers in the first place.

The classification of vacuum gauges measurements to identify possible heating issues during LHC operation has been also considered as a candidate for ML applications. Promising results have been obtained and a Multi-Layer Perceptron has been shown to perform better in terms of recall score. Currently,  more ML and deep learning approaches are under investigation to push further the performance of the classification algorithms. The ultimate goal is to develop an application to be deployed in routine operation during the LHC Run~3.

While the majority of ML applications presented here is connected with beam measurements, the refined analysis of numerical simulations for DA computation has also been considered as a suitable candidate for ML techniques. Indeed, the detection of outliers in the DA values, carried out for sixty realisations of the magnetic errors of the LHC model, has proven to be effectively tackled by Unsupervised Learning. A certain improvement in outlier detection has been achieved by using voting between two different clustering algorithms, but the best outcome has been achieved with a careful post-processing of the results obtained with DBSCAN. Supervised Learning has been attempted too, but even though it has the lowest number of false negatives it creates ten times as much false positives, severely hampering its usability. Thanks to these analyses, it has been possible to study the distribution of outliers for the various realisations of the magnetic field errors as well as for the angle in $x-y$ space. These results need further physical analyses to get more insight in the observed features, but represent a very useful and new tool.

In the future, the time evolution of DA will be considered in conjunction with ML techniques. It is known from theory that such an evolution follows well-defined scaling laws~\cite{Giovannozzi:2018wmm,Giovannozzi:2018igq,VanderVeken:2018sqp,Bazzani:2019csk}. These laws can be used to extrapolate a CPU-intensive simulation, performed over a relatively short number of turns, to realistic timescales in an inexpensive way. So far, this has led to very promising results~\cite{Bazzani:2019csk}, although the fitting procedure is sensitive to several details of the numerical data. Here, a very interesting approach, which is currently being investigated, is to use ML techniques to improve the fitting procedure, e.g.\ by finding a set of optimal fitting weights for the deterministic DA models.
\section*{Acknowledgments}
We would like to thank the LHC team of the Operations Group for the support during the experimental sessions. 

EPFL studies are supported by the Swiss Accelerator Research and Technology institute (CHART).
\bibliographystyle{plain}
\bibliography{references}
\end{document}